\documentclass{aa}  
\usepackage{changes}
\usepackage{graphicx}
\usepackage{xcolor}
\usepackage{gensymb}
\usepackage{txfonts}
\usepackage{indentfirst}
\usepackage{ulem}
\usepackage{mathtools}

\usepackage{hyperref}
\usepackage[version=4]{mhchem}
\usepackage{gensymb}
\usepackage{tikzsymbols}
\usepackage{hypcap}
\graphicspath{{./}{figures/}}

\newcommand{\revisionone}[1]{\textcolor{black}{#1}}

\begin{document}

   \title{Understanding the atmospheric properties and chemical composition of the ultra-hot Jupiter HAT-P-7b}

   \subtitle{II. Mapping the effects of gas kinetics}

    \titlerunning{The ultra-hot atmosphere of HAT-P-7b: II. The effects of kinetic gas composition}

   \author{K. Molaverdikhani \inst{1}
          \and
          Ch. Helling \inst{2,3,4}
          \and
          B. W.P. Lew \inst{5}
          \and
          R.J MacDonald \inst{6,7}
          \and
          D. Samra \inst{2,3}
          \and
          N. Iro \inst{8}
          \and
          P. Woitke \inst{2,3}
          \and
          V. Parmentier\inst{9}
          }

   \institute{Max Planck Institute for Astronomy, Königstuhl 17, 69117 Heidelberg, Germany\\
            \email{Karan@mpia.de}
        \and
            Centre for Exoplanet Science, University of St Andrews, Nort Haugh, St Andrews, KY169SS, UK\\
             \email{ch80@st-andrews.ac.uk,dbss3@st-andrews.ac.uk,oh35@st-andrews.ac.uk, pw31@st-andrews.ac.uk}
         \and
             SUPA, School of Physics \& Astronomy, University of St Andrews, North Haugh, St Andrews, KY169SS, UK
         \and
         SRON Netherlands Institute for Space Research, Sorbonnelaan 2, 3584 CA Utrecht, NL
        \and
            Lunar and Planetary Laboratory, University of Arizona, Tucson, AZ 85721, USA\\
            \email{msteinru@lpl.arizona.edu; wplew@lpl.arizona.edu}
         \and
            Institute of Astronomy, University of Cambridge, Madingley Road, Cambridge, CB3 0HA, UK
         \and
            Department of Astronomy and Carl Sagan Institute, Cornell University, 122 Sciences Drive, Ithaca, NY 14853, USA\\
            \email{rmacdonald@astro.cornell.edu}
        \and
              Institute for Astronomy (IfA), University of Vienna,
              T\"urkenschanzstrasse 17, A-1180 Vienna\\
              \email{nicolas.iro@univie.ac.at}  
        \and
            Department of Physics, University of Oxford, Parks Rd, Oxford, OX1 3PU, UK\\
            \email{vivien.parmentier@physics.ox.ac.uk}
            }
   \date{Received November 2, 2019; accepted January 10, 2020}

 
  \abstract
   {}
   {The atmospheres of ultra-hot Jupiters (UHJs) are commonly considered to be at thermochemical equilibrium. We aim to provide disequilibrium chemistry maps for a global understanding of the chemistry in HAT-P-7b's atmosphere and assess the importance of disequilibrium chemistry on UHJs.}
   {We apply a hierarchical modelling approach utilising 97 1D atmospheric profiles from a 3D General Circulation Model (GCM) of HAT-P-7b. For each atmospheric 1D profile, we evaluate our kinetic cloud formation model consistently with the local gas-phase composition in chemical equilibrium. This serves as input to study the quenching of dominating CHNO-binding molecules. We  evaluate quenching results from a zeroth-order approximation in comparison to a kinetic gas-phase approach.}
   {We find that the zeroth-order approach of estimating quenching points agrees well with the full gas-kinetic modeling results. However, it underestimates the quenching levels by about one order of magnitude at high temperatures. Chemical disequilibrium has the greatest effect on the nightside and morning abundance of species such as H, \ce{H2O}, \ce{CH4}, \ce{CO2}, \ce{HCN}, and all C$_n$H$_m$ molecules; heavier C$_n$H$_m$ molecules are more affected by disequilibrium processes. CO abundance, however, is affected only marginally. While dayside abundances also notably change, those around the evening terminator of HAT-P-7b are the least affected by disequilibrium processes. The latter finding may partially explain the consistency of observed transmission spectra of UHJs with atmospheres in thermochemical equilibrium. Photochemistry only negligibly affects molecular abundances and quenching levels.}
   {In general, the quenching points of HAT-P-7b's atmosphere are at much lower pressures in comparison to the cooler hot-jupiters. We propose \revisionone{several avenues to look for the effect of disequilibrium processes on UHJs that are, in general, based on abundance and opacity measurements at different local times.} It remains a challenge to completely disentangle this from the chemical effects of clouds and that of a primordial non-solar abundance.}


  \maketitle 

\titlerunning{Kinetic gas composition of ultra-hot Jupiter HAT-P-7b}

%

\section{Introduction}
Ultra-hot Jupiters are highly irradiated, tidally locked planets with extreme day/night temperature differences (e.g. \citealt{delrez2017high,Parmentier18,bell2018increased,kreidberg2018global,komacek2018effects,lothringer2018extremely,lothringer2019influence,molaverdikhani2019cold,tan2019atmospheric,arcangeli2019climate,mansfield2019evidence,wong2019exploring}). A day/night temperature difference of $\geq 2000$\;K on, for example,  WASP-18b and HAT-P-7b, results in different chemical regimes. The hot dayside atmosphere is predominately composed of H/He and largely cloud-free, while the cold nightside is composed of a \ce{H2}/He atmosphere where cloud particles can form in abundance (\citealt{2019arXiv190108640H,paperI}).

The extreme insolation received by ultra-hot jupiters, resulting in large atmospheric scale heights, has rendered them prime targets for spectroscopic studies. The most extreme of these planets is KELT-9b; the hottest known exoplanet with an equilibrium temperature of $\sim4000$\;K. High-resolution HARPS-North observations of KELT-9b indicate an abundance of ions, such as \ce{Fe+} and \ce{Ti+}, in its upper atmosphere (\citealt{hoeijmakers2018atomic}). The non-detection of atomic Ti, and less significant detection of Fe, point toward a highly ionised environment. This proliferation of ions is consistent with expectations from thermochemical equilibrium at these temperatures (\citealt{kitzmann2018peculiar}). The somewhat cooler world MASCARA-2b/KELT-20b (T$_{\rm eq}\sim2300$\;K) also shows signs of ions, namely \ce{Ca+}, and \ce{Fe+}, in addition to Na in its transmission spectrum (\citealt{2019arXiv190512491C}). Successes such as these, across a wide temperature regime, demonstrate the observational accessibility of these atmospheric systems.

The progress in detecting spectral features in spectra of ultra-hot Jupiters motivates detailed theoretical modelling of their atmospheres. 1D simulations, e.g., suggest these atmospheres to be in thermochemical equilibrium at the pressures probed by low resolution spectroscopy, i.e. larger than 1\;mbar (\citealt{kitzmann2018peculiar}). However, hydrodynamic transport processes (advection/winds), turbulent eddies and interactions with a radiation field or charged particles (cosmic rays) can affect the local gas phase composition if they favour (or enable) a certain reaction path over its backward/forward reaction. 

It has been shown that turbulent and molecular diffusion, photochemical processes, as well as interaction with free electrons may alter the chemical composition of planetary atmospheres from thermochemical equilibrium to a disequilibrium chemistry state \citep[see e.g.][]{line2010high, moses2011disequilibrium, venot2012chemical, miguel2013exploring, moses2014chemical,2016ApJS..224....9R,2017ApJS..228...20T,helling2019lightning}. While previous studies have been usually conducted on colder planets, \citet{kitzmann2018peculiar} reported that the chemical state of KELT-9b's atmosphere also deviates from thermochemical equilibrium at pressures less than 1\;mbar in a 1D setup. Moreover, according to their simulations, disequilibrium processes drive some of the atomic and molecular abundances away from their chemical equilibrium states at pressures even larger than 1\;mbar. For instance, assuming an inversion in the atmosphere of KELT-9b, they report a pressure of $\sim$0.03\;bar at which or less than that \ce{H2}, \ce{H2O}, and \ce{CO2} abundances deviate from their thermochemical equilibrium values. \ce{CO} and \ce{HCN} abundances also deviate from their thermochemical equilibrium values at around a similar pressure, 0.02\;bar. Transmission spectroscopy is sensitive to these pressures; suggesting that disequilibrium processes might have observable fingerprints on the chemical composition of highly irradiated planets as well.

The 3D nature of planets and their resulted colder nightside could also enhance the effect of disequilibrium processes by extending the quenching level of major opacity species to larger pressures at the terminators and nightside. Consequently, 3D studies are needed to take into account local processes (such as disequilibrium chemistry) and connect the probed regions of the atmosphere to its global properties. Here, we begin to assess the nature of disequilibrium processes in ultra-hot Jupiters using full 3D modelling of the governing chemical and physical processes. The chosen target of our investigation is the ultra-hot Jupiter HAT-P-7b (T$_{\rm eq}\sim2100$\;K). We began this exercise by studying global cloud formation, and its impact on local elemental abundances, in \citealt{paperI} (hereafter, Paper I). We continue our investigation by addressing the role of kinetic processes in altering the 3D gas-phase composition from thermochemical equilibrium expectations.


We note, however, that models linking kinetic gas-phase chemistry to dynamic atmosphere simulations are quite time-consuming. Usually, therefore, relaxation schemes are applied to omit the long computational times required to assure independence of the solution from the initial conditions, \citep[see e.g.][]{cooper2006dynamics,drummond2018b,2018ApJ...869...28D,mendonca2018b}. In such schemes, it is assumed that if the timescale for hydrodynamic transport processes becomes longer than the chemical timescale, then the local atmosphere will achieve chemical equilibrium (wherein both forward and backward reactions occur with a similar rate). This mostly occurs at hot and high-pressure atmospheric levels or in hydrodynamically inactive regions. The atmosphere will be driven out of thermochemical equilibrium if the hydrodynamic timescales (also called ``mixing timescales'')  become shorter than the chemical timescales. This occurs, for example, if a cell of material is transported more rapidly than its thermodynamic state can adjust to its new surrounding \citep[e.g.][]{prinn1977carbon,griffith1999,2006ApJ...647..552S,cooper2006dynamics,2018ApJ...869...28D}. Hence, the composition of the cell may remain that of a hotter gas as long as there is no decay reaction fast enough (e.g. \ce{N2}/\ce{NH3}, \ce{CO}/\ce{CH4}). This had been demonstrated for brown dwarfs, e.g., by \citet{saumon2000molecular}. The effect of such a dynamically-induced gas disequilibrium on phase curves has been demonstrated, e.g., in \cite{2018arXiv180802011S}. For instance, in wavelength bands with significant methane absorption, the amplitude of phase curves can be dramatically reduced. This effect is expected to be more significant for planets on which the nightside becomes cold enough that, under equilibrium chemistry, methane becomes the dominant nightside carbon-bearing species while CO prevails on the dayside \citep{2018arXiv180802011S}. We, however, note that \ce{CH4} maybe destroyed by lightning locally \cite{2014ApJ...784...43B}.

Transport induced chemical disequilibrium mostly occurs for gaseous species at higher-altitude atmospheric conditions. Here, it may lead to a chemical stratification as each species reacts according to its own timescale, and each species diffuses on its own time scale, in principle. The pressure level at which this transition between thermodynamic equilibrium and kinetic disequilibrium takes place is called the {\it quenching point} or quenching level \citep{smith1998estimation}. The determination of the quenching level varies between models which, for example, apply a constant diffusion constant for all (gas and cloud) species \citep{ZhangShowman2018a,ZhangShowman2018b}. We further note that quenching levels can only be studied for molecules with known kinetic forward and backward rates, and with known diffusion constants.

The relative gas number densities, i.e. mixing ratios, of the affected species at pressures lower than the pressure of the quench point (i.e. higher altitudes) are believed to remain equal to the equilibrium abundances at the quench point, or so called quenched-abundances \citep[see e.g.][]{prinn1977carbon,moses2011disequilibrium,venot2018better}. This uniform (``quenched'') mixing ratio profile has been frequently used in the atmospheric retrievals as a simplified approach to increase the computational speed \citep[see e.g.][]{madhusudhan2011high, tsiaras2018population, pinhas2018h2o}. We propose a fast zeroth-order method (based on \citet{venot2018better} results) to estimate the quench levels and compare it with the gas-phase kinetic estimations.

In addition, under different circumstances and for different species, the abundance above the quenched level can deviate from a constant value in a chemical kinetic context \citep[see e.g.][]{molaverdikhani2019cold2}. If significant, this deviation might pose a problem in the retrieved abundances as shown by, e.g., \citet{changeat2019towards}. We discuss the validity of constant mixing ratio approximation for the case of HAT-P-7b, in \hyperref[sec:kineticsRes]{Section}\;\ref{sec:kineticsRes}.

We, therefore, aim to quantitatively assess the importance of kinetic gas-phase chemical processes that are driven by hydrodynamic motion or by photochemistry, and map them in the case of HAT-P-7b; an ultra-hot Jupiter. We describe our hierarchical modelling approach in Section \ref{s:ap} and present the results on the quenching levels for the HAT-P-7b's atmosphere in Section \ref{s:QuenLev} where we add a critical evaluation of the ``quenching level'' concept. In Section \ref{s:maps} we highlight our findings on how much the abundances of \ce{H2O}, CO and \ce{CH4} change due to diffusion (both turbulent eddies and molecular) and photochemistry. We summarize and conclude our findings in Sections \ref{s:discussion} and \ref{s:conclusions}.


\section{Approach}\label{s:ap}
We use our modelling tools as virtual laboratories, expanding our hierarchical approach of modelling environments from Paper I. In Paper I, we utilised 97 1D profiles extracted from the 3D GCM calculated for the cloud-free HAT-P-7b \citep[see Fig.1 in ][]{mansfield2018hst}. \revisionone{Latitudes and longitudes are distributed with 22.5$\degree$ and 15$\degree$ intervals respectively; resulting in 96 1D profiles in one hemisphere. \revisionone{The zero longitude reference is defined at the substellar point.} An additional node at longitude=90$\degree$ and latitude=85.5$\degree$ was also chosen to probe the polar region.} We then applied our kinetic cloud formation model to solve for the gas phase composition under chemical equilibrium, while consistently accounting for element abundance changes due to cloud formation. The next modelling step is to use the 97 1D (T$_{\rm gas}$(z), p$_{\rm gas}$(z), $w(x,y,z)$)-profiles for HAT-P-7b (Fig.~1 in Paper I: T$_{\rm gas}$(z) is the local gas temperature [K], p$_{\rm gas}$(z) is the local gas pressure [bar],  $w$(x,y,z) is the local vertical velocity component [cm s$^{-1}$]\revisionone{; all are direct outputs of the GCM}) to derive the diffusion coefficient K\textsubscript{zz}, which assumes that all gas species have the same eddy diffusion coefficient.

In this section, we outline the two approaches we adopt to study the effect of gas-phase kinetic processes on the atmosphere of HAT-P-7b. We follow a hierarchical approach, wherein we begin in Section~\ref{sec:zeroth_appx} with a simple quenching level calculation (i.e. `zeroth-order'), before increasing the complexity in Section~\ref{sec:kineticmodelling} by considering full gas-phase chemical kinetic modelling.

\subsection{Quenching levels: a zeroth-order approximation} \label{sec:zeroth_appx}

\begin{figure}
\includegraphics[width=\columnwidth]{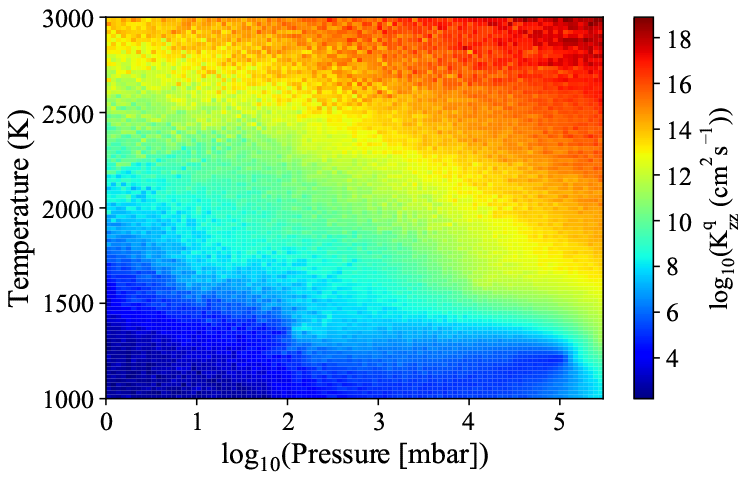}
\caption{Estimation of eddy coefficient needed for quenching to happen as a function of pressure and temperature\revisionone{, K$^{q}_{zz}$(p,T). \citet{venot2018better} report the quenching pressures for several pairs of temperature-vertical mixing strength (in their Figure~1), and we interpolate their results to obtain a continuous K$^{q}_{zz}$ function at any chosen pressure and temperature. See Sec.~\ref{sec:zeroth_appx} for more details.}}
\label{fig:quench interpolation}
\end{figure}

Modelling kinetic gas chemistry is computationally expensive. Hence, it can be instructive to explore fast, but reasonably accurate, approaches to investigate chemical quenching in 3D atmospheric models. As a zeroth-order approximation, we adapt the \revisionone{calculated quenching pressures by} \citet{venot2018better} \revisionone{(presented in their Figure 1)} to find the quenching levels for each 1D trajectory in the 3D model of HAT-P-7b. In brief, \citet{venot2018better} use a series of different thermal profiles (following analytical models from \citet{parmentier2014non}) to determine the quenching level as a function of temperature and eddy diffusion coefficient (K\textsubscript{zz}) by examining the criteria of at least one of their main species (i.e. \ce{H2}, H, \ce{H2O}, \ce{CH4}, CO, \ce{CO2}, \ce{N2}, \ce{NH3}, HCN, \ce{CH3}, and OH) deviating from its thermochemical equilibrium abundance.

Given the monotonic nature of these functions (the quenching pressure as a function of temperature and K$_{zz}$), their inverse functions can be easily constructed, i.e. eddy diffusion coefficients at the quenching points (K$^{q}_{zz}$) as a function of pressure and temperature. We linearly interpolate/extrapolate these K$^{q}_{zz}$ values (using radial basis functions) in the range of 1000\;K$<$T$_{gas}<$3000\;K and 1\;mbar$<$P$_{gas}<$10$^6$\;mbar; see \hyperref[fig:quench interpolation]{Fig.}\;\ref{fig:quench interpolation}. Comparing K$^{q}_{zz}$ with the derived eddy diffusion coefficient at any given pressure and temperature from the GCM's 1D trajectories provides an estimate for whether that atmospheric level is quenched or not (i.e. if K$_{zz}(P,T)>$K$^{q}_{zz}(P,T)$, it is quenched).

\begin{figure}
\includegraphics[width=\columnwidth]{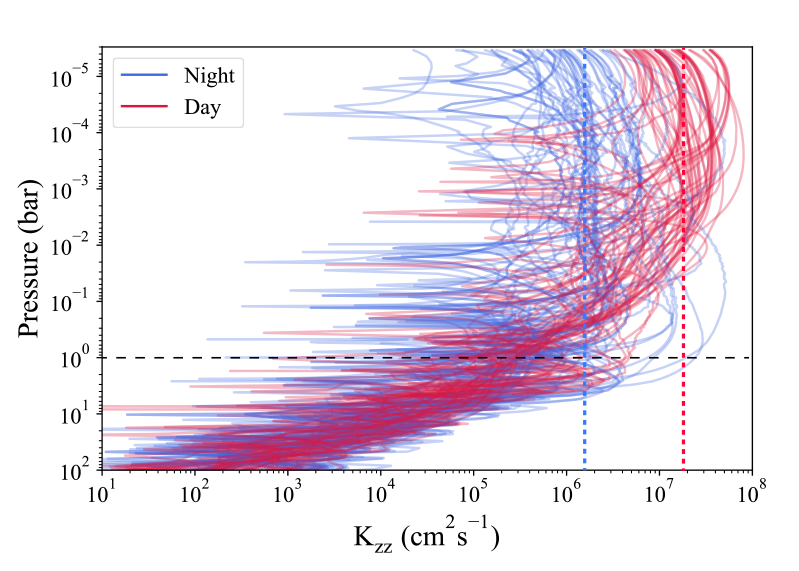}
\caption{Calculated eddy diffusion coefficient (K$_{zz}$) profiles using Equation~\ref{eq:eq_Kzz}. The average eddy diffusion coefficient at pressure larger than 1\;bar roughly follows a simple relation as \revisionone{log$_{10}$(K$_{zz}$)=5.5-1.7log$_{10}$(p\textsubscript{gas})}. Above this pressure and on the dayside, K$_{zz}$ has an average values around 2$\times10^7 cm^2s^{-2}$ (vertical red dashed line), while it is weaker by about one order of magnitude on the nightside (vertical blue dashed line).}
\label{fig:Kzz}
\end{figure}

In 1D kinetic models, the eddy diffusion coefficient (K$_{zz}$) is usually estimated by multiplying the vertical wind speed ($w$) by the relevant vertical length scale, e.g. scale height ($H$) (see e.g. \citealt{lewis2010atmospheric,moses2011disequilibrium}). However, \citet{parmentier20133d} derived K$_{zz}$ values from passive tracers in their GCM and found that these values were about two orders of magnitude lower than the traditional approach. Therefore, we approximate the mixing strength as follows:
\begin{equation}
K_{zz}=|w|\cdot H\times 10^{-2}.
\label{eq:eq_Kzz}
\end{equation}
where H, the atmospheric scale height, is given by H=RT/$\mu$g at any pressure level ($R$ is the universal gas constant, $T_{\rm gas}$ is the gas temperature, $\mu$ is the mean molar mass -- defined as the mass of the atmosphere at each level divided by the amount of substance in that level, and $g$ is the gravitational acceleration). The average eddy diffusion coefficient at pressures larger than 1\;bar shows a somewhat linear relation with pressure as log(K$_{zz}$) = 5.5\;-\;1.7log(P$_{gas}$); \hyperref[fig:Kzz]{Fig.}\;\ref{fig:Kzz}. Above this region, the eddy diffusion coefficient is less variable on average; revolving around values of 2$\times10^7$cm$^2$s$^{-2}$ and 2$\times10^6 $cm$^2$s$^{-2}$ on the dayside and nightside, respectively. The stronger K$_{zz}$ on the dayside, along with the cooler atmosphere on the nightside, hint towards the existence of a region between the day and night with a minimum quenching pressure. This is discussed in the results section (Sec.~\ref{sec:zeroth_appx_res}). Note that by taking this approach, small scale and local variations of K$_{zz}$ might be introduced in comparison to the case of a smoothed parametrized K$_{zz}$ approach. Therefore, small scale variations of abundances are possibly not very realistic and the results should be taken for their order-of-magnitude significance.

Although this approach is quite simple, it provides a reasonable overall picture about where we should expect chemical quenching to occur. Nevertheless, a more sophisticated approach should be considered in order to quantitatively assess the importance of disequilibrium processes on Ultra-hot Jupiters. For this, we turn now to chemical kinetic modelling.

\begin{figure*}[!h]
\includegraphics[width=\textwidth]{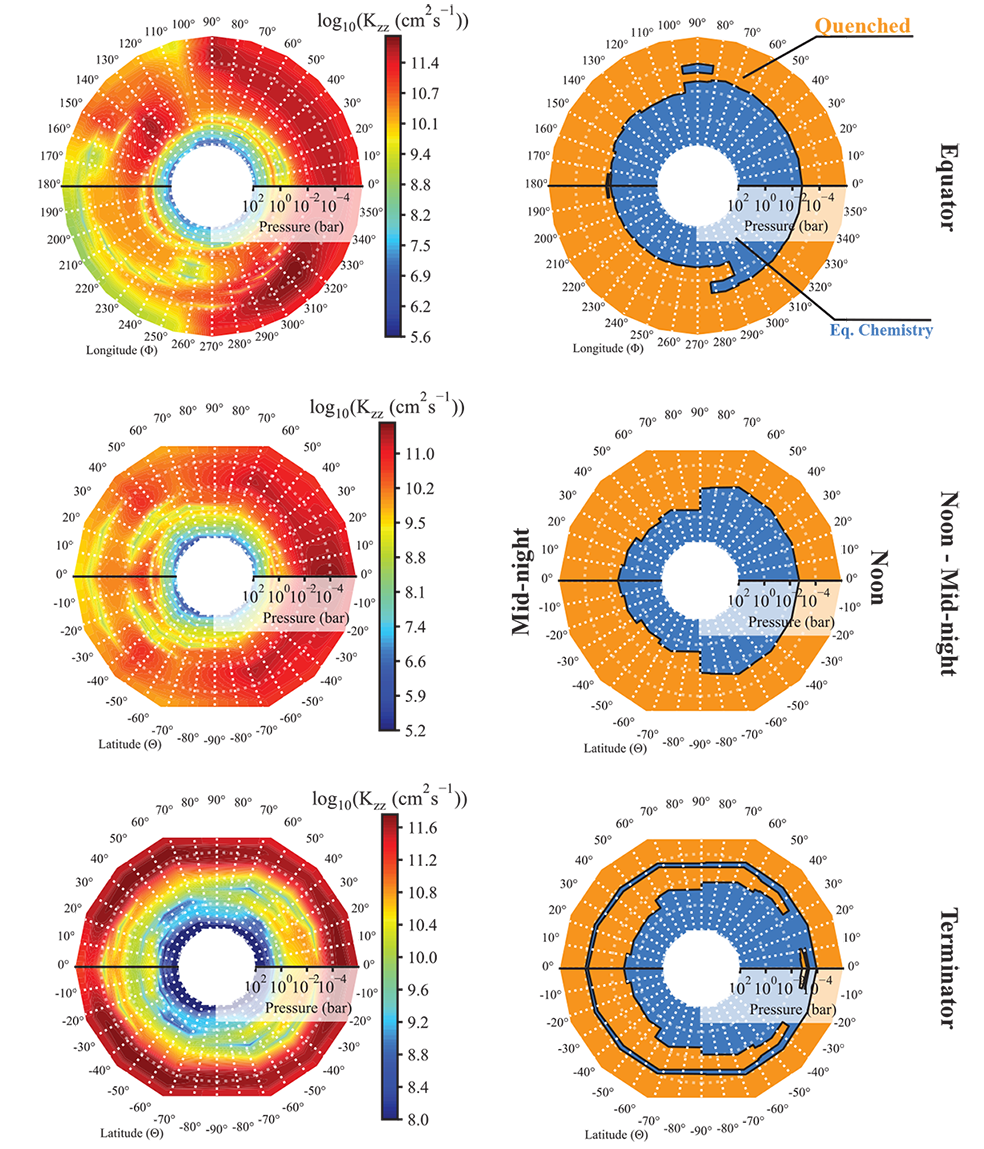}
\caption{{\bf Left)} Calculated K\textsubscript{zz} using Equation~\ref{eq:eq_Kzz}, and {\bf Right)} Calculated quenching levels by following a zeroth order approach. In the blue regions the atmosphere is in thermochemical equilibrium while orange highlights the regions that are expected to be affected by disequilibrium processes and at least one major species deviates from its thermochemical equilibrium abundance.}
\label{fig:quench0thKzzPq}
\end{figure*}

\subsection{Chemical kinetic modelling}
\label{sec:kineticmodelling}

A chemical network is essential to quantitatively trace the chemical abundances resulting from interactions between major and minor species at different pressures and temperatures. For this, we use \texttt{ChemKM}, a photochemical kinetic model \citep{molaverdikhani2019cold2} to perform a detailed study of gas-phase chemical kinetics and quenching levels on HAT-P-7b for CNOH-binding species. \texttt{ChemKM} is a modular package that calculates the vertical distribution and temporal evolution of atmospheric constituents in 1D. It employs the ODEPACK library \citep{hindmarsh1983odepack} and solves the coupled mass-continuity equations as a function of pressure for each species:
\begin{equation}
\frac{\partial n_i}{\partial t} + \frac{\partial \Phi_i}{\partial z} = P_i - n_iL_i,
\label{eq:masscont}
\end{equation}
where n$_i$ is the number density ($\rm cm^{-3}$), $\Phi_i$ is the vertical flux (cm$^{-2}$ s$^{-1}$), P$_i$ is the production rate (cm$^{-3}$ s$^{-1}$), t is time, z is the vertical distance from the center of planet, and L$_i$ is the chemical loss rate (s$^{-1}$) of species i.

Photochemical processes by UV and X-rays, photodissociation by galactic cosmic rays, flux attenuation due to Rayleigh scattering, condensation, ablation, outgassing, molecular diffusion and eddy mixing could all driven the chemical composition of atmosphere out of thermochemical equilibrium. However, here we only explore the effects of diffusion/mixing and photochemical processes by UV and X-ray as the other processes are poorly known for the exoplanets.

In order to solve Eq.~\ref{eq:masscont}, we adopt a reduced version of the \citet{venot2012chemical} network that accounts for most of the dominant species expected at HAT-P-7b's atmospheric conditions. \revisionone{While 97 1D profiles is not a huge number, developing a reduced network optimised for a particular case would reduce the computational time needed for studying the case. We developed the reduced network through a sensitivity analysis of the abundances of major opacity species (mainly water; see Sec.~\ref{ss:gas_opa}). Excluding the species in \citet{venot2012chemical} one by one and calculating the differences between the abundances, resulted in choosing a list of 38 species (and their corresponding reactions). The current reduced network results in less than 10\% differences at all pressures for major species comparing with \citet{venot2012chemical}. This is much less than other uncertainties in the kinetic chemical simulations and justifies order-of-magnitude studies by using this network, which is the aim of this work. Although this approach does not assure the generality of this reduced network for any other environment, it is possible to use it for similar cases to HAT-P-7b, in terms of temperature range, as well as metallicity and C/O ratio (both are assumed to have solar values). Comparing our network with the recently published reduced network by \citet{venot2019reduced} (with 30 species) shows that these networks share 23 species and only differ on some of the heavier molecules. A thorough comparison of these networks demands a further analysis.}

Our reduced network contains 38 species (namely, \ce{H2}, \ce{He}, \ce{H}, \ce{CO}, \ce{H2O}, \ce{CH4}, \ce{N2}, \ce{NH3}, \ce{CO2}, \ce{HCN}, \ce{C2H2}, \ce{O2}, O(3P), O(1D), \ce{C}, N(4S), N(2D), \ce{OH}, \ce{CH3}, $^3$\ce{CH2}, $^1$\ce{CH2}, \ce{CH}, \ce{NH2}, \ce{NH}, \ce{H2CO}, \ce{HCO}, \ce{CN}, \ce{C2H}, \ce{NO}, \ce{HNO}, \ce{NCO}, \ce{HNCO}, \ce{HOCN}, \ce{HCNO}, \ce{C2H3}, \ce{C2H4}, \ce{C2H5}, and \ce{C2H6}), corresponding to 611 kinetic reactions. The UV absorption cross-sections and branching yields were adapted from an updated version of the \citet{hebrard2012neutral} network, which includes 29 photo-reactions for the species in our reduced network. The photo-species are \ce{H2O}, \ce{CO2}, \ce{H2CO}, \ce{OH}, \ce{CO}, \ce{H2}, \ce{CH4}, \ce{CH3}, \ce{C2H2}, \ce{C2H3}, \ce{C2H4}, \ce{C2H6}, \ce{N2}, \ce{HCN}, \ce{NH3}, \ce{NO}, \ce{HCO}.

The irradiation field was determined by interpolating between Phoenix stellar atmosphere models \citep{hauschildt1999nextgen,husser2013new}\revisionone{\footnote{Phoenix models are accessible under http://phoenix.astro.physik.uni- goettingen.de/}}, assuming the following stellar parameters: T$_{*}=$6441\;K, log(g)$=$4.02\;cm s$^{-2}$, and [Fe/H]$=$0.15 \citep{torres2012improved}. While some UV observations by HST/STIS/COS are available for similar stars, such as HD\;27808 \citep[e.g.][]{bowyer1994first}, we use the Phoenix models to take a coherent approach. This means no data stitching is required in order to construct the host star's flux between 1\;nm to 300\;nm (the wavelengths at which the irradiation becomes relevant to the photolysis of atmospheric constituents). In addition, the dayside of the planet is very hot and, hence, likely to be insensitive to the photochemistry (see Sections \ref{sec:photochem} for details). Therefore, small differences in the UV portion of the constructed stellar spectra using different methods (i.e. data stitching or Phoenix model) is not expected to cause a noticeable change in the results, and the Phoenix model provides an adequately accurate estimation of HAT-P-7's spectrum. \revisionone{It is also worth noting that small differences in the stellar spectra are inconsequential due to relatively large uncertainties in the data used to model photodissociation. These data, such as cross-sections and branching ratios, are poorly known at temperatures higher than 300\;K \citep{venot2018vuv}.}


Following our hierarchical modelling approach, we use the gas-phase chemical equilibrium results from Paper I as the input values. We then compute the thermochemical abundances of species considered in our kinetic network to calculate the initial abundances for the kinetic gas-phase modelling in this paper, as discussed in Sect. ~\ref{sec:kineticmodelling}. The results from Paper I include the effect of cloud formation on the element abundances through depletion and/or enrichment (see Sect 2.1. in Paper I). We  use the same 1D  atmospheric temperature-pressure profiles (Figs. 1 and 3 in Paper I). We present and discuss the results of our kinetic gas-chemical models in \hyperref[sec:kineticsRes]{Section}\;\ref{sec:kineticsRes}.

In order to solve the continuity equation, Eq. \ref{eq:masscont}, we follow \cite{2012ApJ...761..166H} (Eq. 2) and express the vertical transport flux of species $i$ ($\Phi_i$) as follow:
\begin{equation}
\Phi_i=-K_{zz}n\frac{\partial f_i}{\partial z}-D_in\left(\frac{\partial f_i}{\partial z}+\frac{f_i}{H_i}-\frac{f_i}{H_0}+\frac{\alpha_i}{T}\frac{dT}{dz}f_i\right) 
\label{eq:vertical_trans}
\end{equation}
where K$_{zz}$ is approximated using Eq.~\ref{eq:eq_Kzz}, $n$ is the total number density, $H_0$ is the mean scale height of the atmosphere, $H$ is the molecular scale height, $T_{\rm gas}$ is the gas temperature, $f_i$ is mixing ratio, $D_i$ is molecular diffusion coefficient, and $\alpha_i$ is the thermal diffusion factor of species $i$.

The thermal diffusion factor is usually set to zero, except for light species such as H, \ce{H2}, and He, that can be estimated with -0.25 \citep[e.g.][]{marshall2013theory}. To estimate molecular diffusion coefficients, $D_i$, we use the Chapman and Enskog equation \citep{poling_properties_2000}: 
\begin{equation}
D=\frac{3}{16}\frac{(4\pi kT/M_{AB})^{1/2}}{n\pi \sigma^2_{AB}\Omega_{D}}f_D
\label{eq:diff}
\end{equation}
where $M_{AB}$ is defined as $2[(1/M_A)+(1/M_B)]^{-1}$ ($M_A$, $M_B$ being the molecular weights of species A and B, respectively), $\Omega_{D}$ is the collision integral for diffusion, $\sigma_{AB}$ is the characteristic length of the intermolecular force law, $f_D$ is a correction term (usually of order unity), n is number density of molecules in the mixture, k is the Boltzmann’s constant, and T is the gas temperature of the mixture. A detail description of how we approximate the molecular diffusion coefficients is provided in Appendix \ref{s:appendix_mol_diff}.

As pointed out, the gas phase number density profiles of the 38 CHNO-bearing species, as result the kinetic cloud formation presented in Paper-I, are used as input values for our chemical kinetic gas-phase calculations. These input molecular abundances already account for elemental depletion or enrichment by cloud formation processes. However, we note here that the comparison with our gas-chemistry results from Paper I bears a conceptional problem, namely the definition of an interface.

\paragraph{Defining the interface:} 
A challenge for hierarchical models may be to define an interface that is unambiguous and practical at the same time.  The comparison between results from an equilibrium gas-phase code with a kinetic rate network gas-phase code is challenged by the kinetic model only treating a limited number of gaseous species. In this paper, the interface is therefore defined by the number densities, $n_{\rm x}/n_{\rm tot}$ [cm$^{-3}$], of those species that can be considered by the kinetic code (\ce{H2}, \ce{He}, \ce{H}, \ce{CO}, \ce{H2O}, \ce{CH4}, \ce{N2}, \ce{NH3}, \ce{CO2}, \ce{HCN}, \ce{C2H2}, \ce{O2}, O(3P), O(1D), \ce{C}, N(4S), N(2D), \ce{OH}, \ce{CH3}, $^3$\ce{CH2}, $^1$\ce{CH2}, \ce{CH}, \ce{NH2}, \ce{NH}, \ce{H2CO}, \ce{HCO}, \ce{CN}, \ce{C2H}, \ce{NO}, \ce{HNO}, \ce{NCO}, \ce{HNCO}, \ce{HOCN}, \ce{HCNO}, \ce{C2H3}, \ce{C2H4}, \ce{C2H5}, and \ce{C2H6}). Hence, the \textbf{\textit{input}} (interface) values for all quenching approaches (Sect.~\ref{sec:zeroth_appx}-\ref{sec:kineticmodelling}) considered here are set by the results from our thermochemical equilibrium calculation in Paper I. The \textbf{\textit{initial}} abundances in our kinetic model are then computed by calculating the new thermochemical equilibrium abundances of these 38 species, given these input values. Any significant deviation from these initial abundances \revisionone{through time-dependent kinetic calculations} is considered as an indication of the fingerprints of disequilibrium processes, as discussed in Sect.~\ref{sec:kineticsRes}.

This choice of interface, however, results in lower element abundances of C, N and O if they are derived based on the above listed 38 species only (rather than from the full list of species used in Paper I). If the element abundances were kept consistent between the two modelling steps, we would have inconsistencies in the initial molecular number densities compared to the values derived in Paper I. See Sec.~\ref{ss:CtoO} for more details. 

We also note that our 1D approach neglects horizontal transport, which could alter the exact location of the quenching and abundances \citep[e.g.,][]{agundez2014,drummond2018b,mendonca2018b}. This limitation is discussed in Section \ref{s:discussion}.

\section{Quenching levels} \label{s:QuenLev}

\subsection{Results of zeroth-order approximation} \label{sec:zeroth_appx_res}

The quenching levels by the zeroth-order approximation are determined by following the criteria of at least one of  \citet{venot2018better} model's main species (i.e. \ce{H2}, H, \ce{H2O}, \ce{CH4}, CO, \ce{CO2}, \ce{N2}, \ce{NH3}, HCN, \ce{CH3}, and OH) deviating away from its thermochemical equilibrium abundance. Therefore, this approximation does not associate the quenching levels to any specific species, unlike the kinetic approach.

\hyperref[fig:quench0thKzzPq]{Figure}\;\ref{fig:quench0thKzzPq} top panels reproduce the transection maps of K$_{zz}$ and the quenched regions at the equator. Despite weaker vertical mixing at the night-side, quenching levels are deeper ($p_q^{\rm night}\approx$10-100\;mbar) relative to the dayside ($p_q^{\rm day}\approx$1-10\;mbar) due to the much hotter atmospheric environment by about 2000\;K on the dayside. This is more evident in the noon-midnight maps, i.e. \hyperref[fig:quench0thKzzPq]{Fig.}\;\ref{fig:quench0thKzzPq} middle-panels. \revisionone{The noon-midnight transection map shows a strong increase in the quenching pressure from the dayside ($\sim$1\;mbar) to the nightside ($\sim$100\;mbar), with no significant latitudinal quench point variation on either sides and a sudden change in the quenching levels over the pole.} The transition, however, is more gradual between the evening and morning terminators, see \hyperref[fig:quench0thKzzPq]{Fig.}\;\ref{fig:quench0thKzzPq} bottom-panels.

\begin{figure}
\includegraphics[width=\columnwidth]{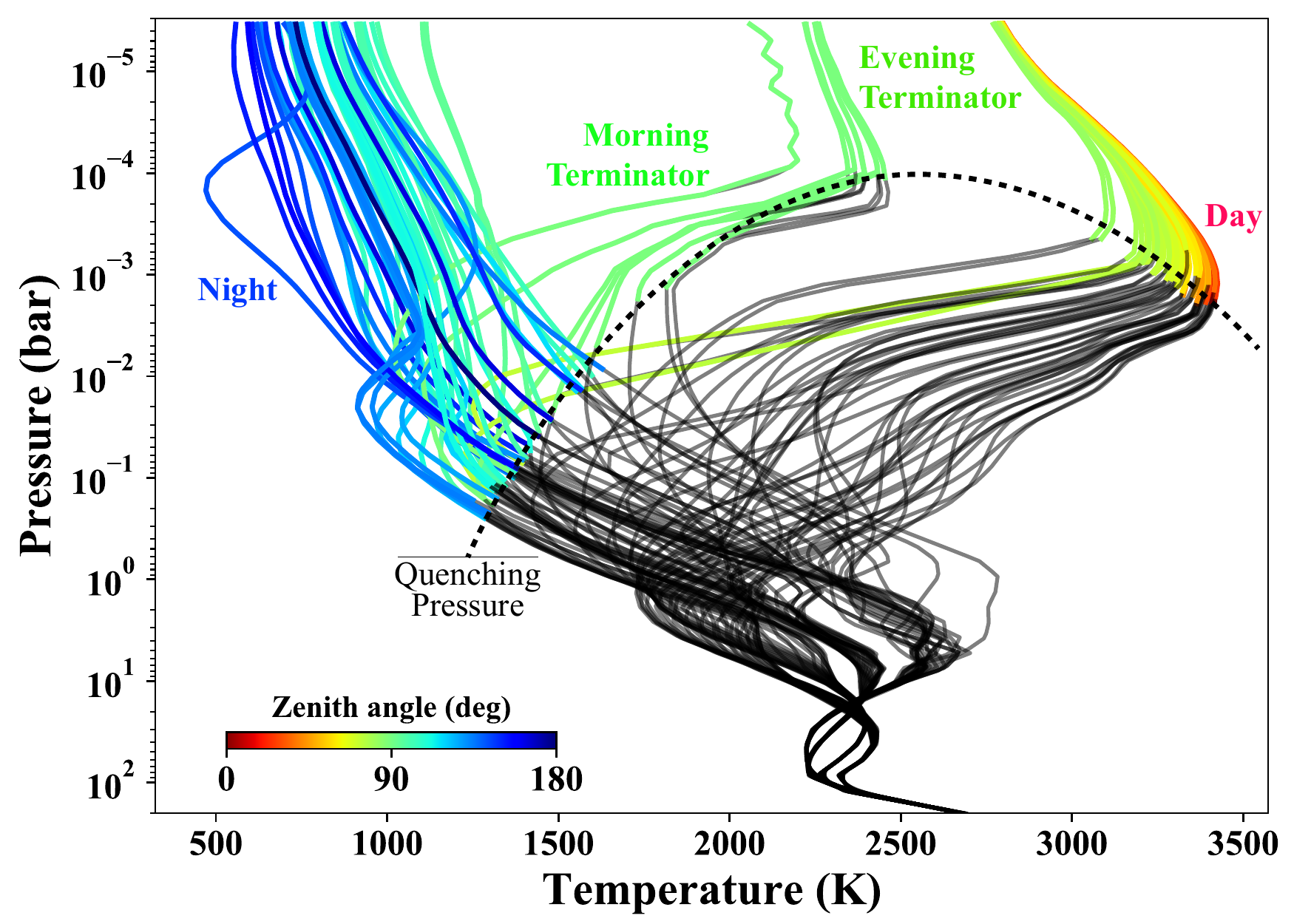}
\caption{Variation of quenching pressure with the zenith angle, using the zeroth-order approximation (Sec.~\ref{sec:zeroth_appx_res}). Temperature profiles of all trajectories are color-coded above the quenching pressures. Minimum quenching pressure occurs at the evening terminator; where the atmosphere contributes significantly in the transmission spectra. Hence, the effects of disequilibrium chemistry is expected to be less pronounced in the transmission spectra. \revisionone{Variation of quenching pressure is represented with the dashed line.}}
\label{fig:qp0th}
\end{figure}

The value of the quenching pressure depends on the mixing efficiency (K$_{\rm zz}$) and temperature. \citet{Komacek2019verticalmixing} found that in general K$_{\rm zz}$ increases with temperature in the mbar regime. As both of these quantities decrease at higher zenith angles (i.e. toward the nightside), there should be a region where the quenching pressure is at its minimum values (i.e. quenching point occurs at its maximum altitude) for the affected species by \revisionone{diffusion}. This is evident in \hyperref[fig:qp0th]{Figure}\;\ref{fig:qp0th} where the quenching levels have a maximum altitude at the evening (hotter) terminator according to our zeroth-order approximation. Hence, thermochemical equilibrium assumption is the most relevant at the evening terminator. This may partially support the applicability of this assumption for the transmission spectroscopy of ultra hot Jupiters. Further analysis will be performed in a forthcoming paper to investigate this in more details.

As shown in both \hyperref[fig:quench0thKzzPq]{Fig.}\;\ref{fig:quench0thKzzPq} and \ref{fig:qp0th}, deeper regions usually remain in chemical equilibrium, however, there might be some detached regions at the lower pressures which do not satisfy the \citet{venot2018better} quenching criteria. For instance, the blue ring in the terminator transection map, \hyperref[fig:quench0thKzzPq]{Fig.}\;\ref{fig:quench0thKzzPq} bottom-right panel, where the chemistry departs from the quenched profiles and returns toward the equilibrium profiles at some pressures. This is mostly due to a combination of temperature inversion and lower vertical velocity. Nevertheless, if only one quenching level at each 1D trajectory is desired, then the deepest level at which the quenching occurs can be selected as the quenching point.

\begin{figure*}[h]
\includegraphics[width=\textwidth]{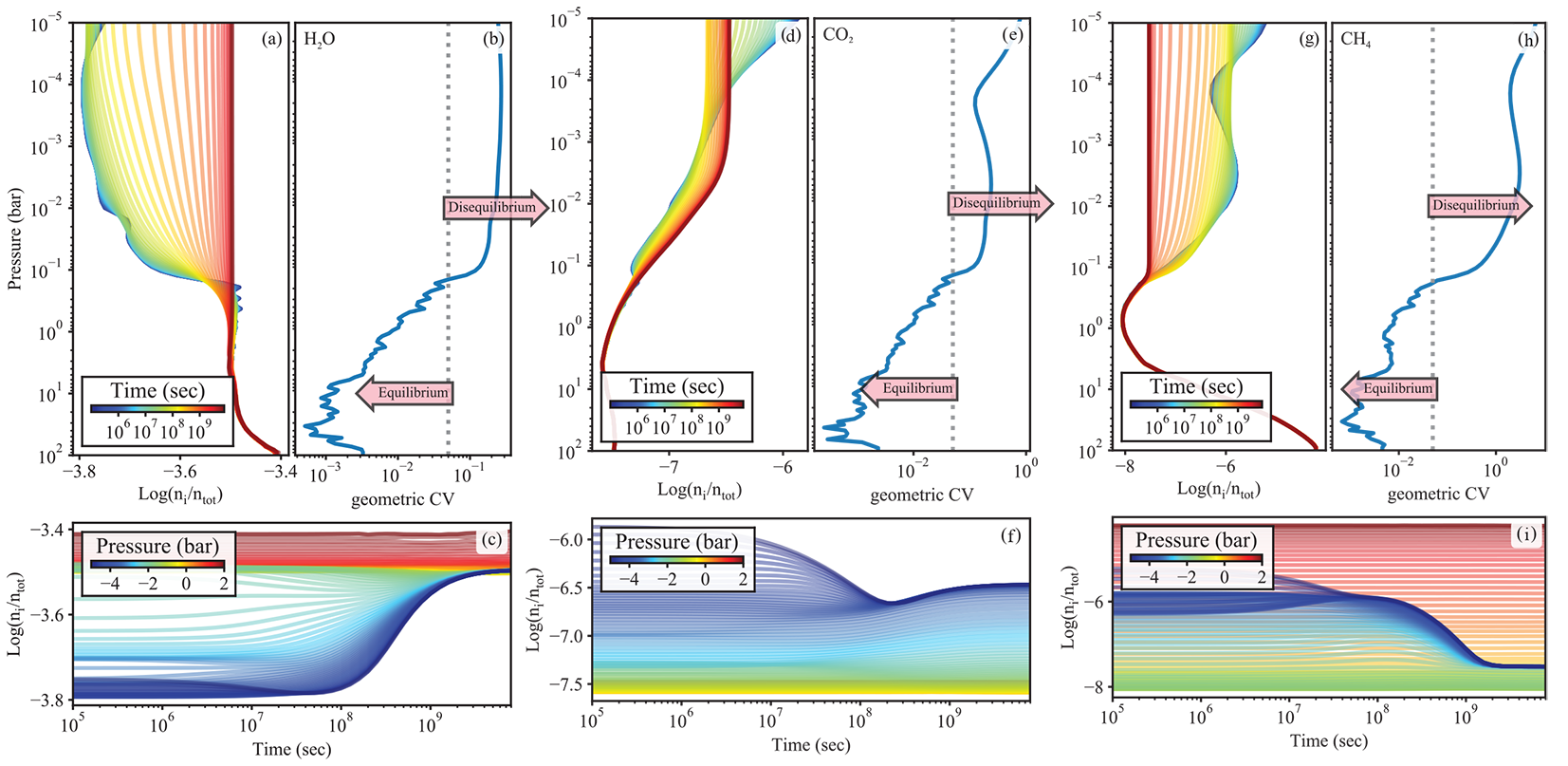}
\caption{(a, d, g) Temporal variation of \ce{H2O}, \ce{CO2}, and \ce{CH4} vertical abundances at the equator and longitude=225$\degree$, i.e. night-side. (b, e, h) Geometric Coefficient of Variation (gCV) of abundances\revisionone{, from initial values to a steady-state at $\sim5\times10^9$\;sec. A} higher value represents a stronger variation of abundance at any given pressure level. A choice of gCV=0.05 seems to reasonably separate the chemical equilibrium and disequilibrium regions (dashed gray vertical lines). (c, f, i) The quenching timescales. An invariant abundance profile at the longer times ensures a new steady-state. See \hyperref[sec:kineticsRes]{Section}\;\ref{sec:kineticsRes} for more details. We note that the quenching level cannot be associated with one particular pressure for all species, but usually spans more than 1 order of magnitude in pressure.}
\label{fig:gCV profiles}
\end{figure*}

\begin{figure*}[h]
\includegraphics[width=\textwidth]{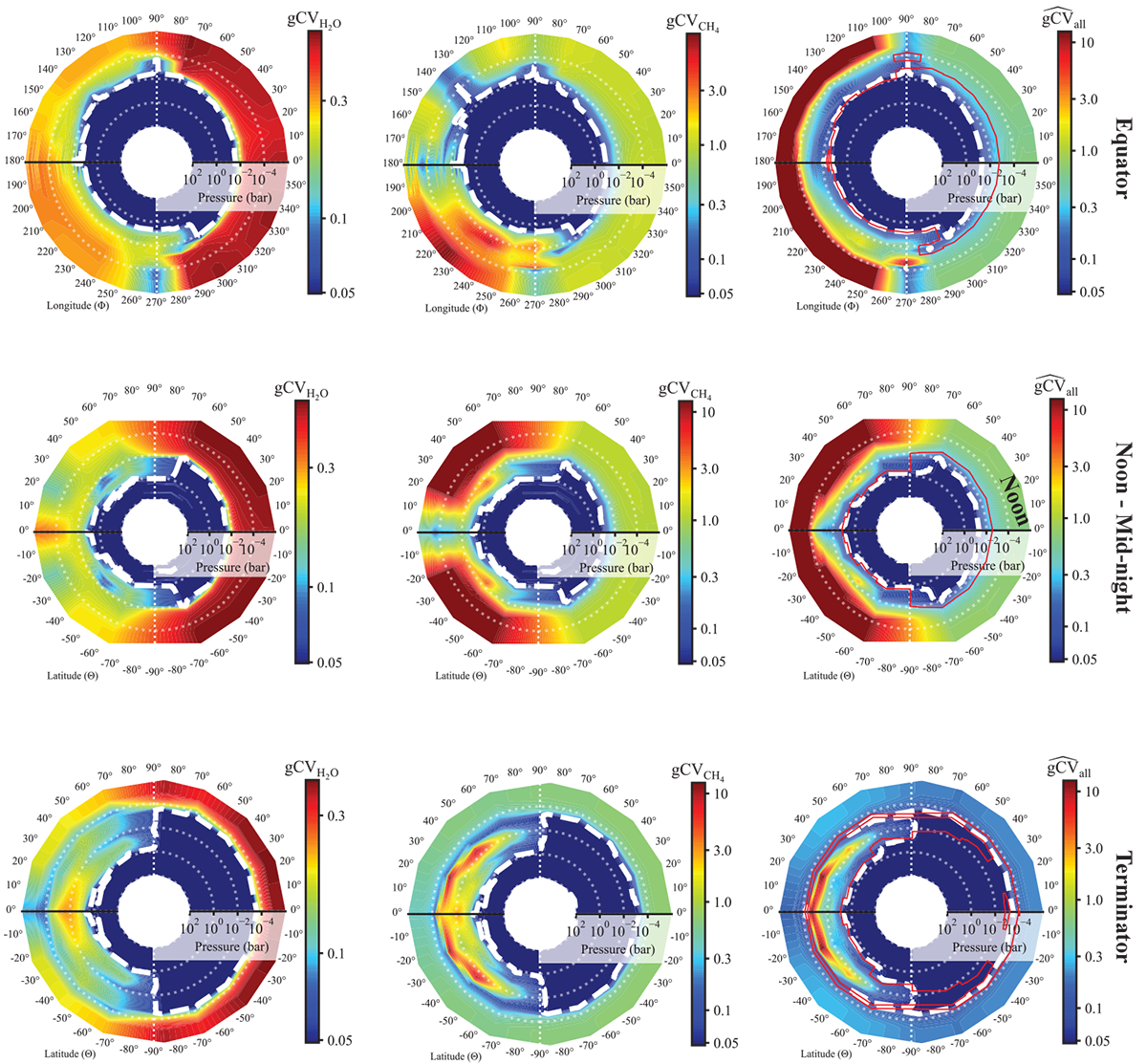}
\caption{Estimation of quenching levels using the chemical kinetic modeling (\hyperref[sec:kineticsRes]{Section}\;\ref{sec:kineticsRes}). Top panels) Transection maps of the geometric Coefficient of Variance (gCV) of \ce{H2O}, \ce{CH4}, and all species abundances at the equator (Latitude=0$\degree$). We choose gCV=0.05 to separate the chemical equilibrium and disequilibrium regions (dashed white lines). This boundary is in good agreement with the zeroth-order approximation results (red solid lines in the right panels; \hyperref[fig:quench0th maps]{Fig.}\;\ref{fig:quench0th maps}). Middle panels) Similar to the top panels but at the noon-midnight plane (Longitude=0$\degree$ or 180$\degree$). Bottom panels) Similar to the top panels but at the terminator (Longitude=90$\degree$ and 270$\degree$). $\widehat{\rm gCV}_{\rm all}$ map suggests a morning-evening asymmetry where the dawn-side abundances have been driven more significantly out of thermochemical equilibrium and at deeper pressure levels.}
\label{fig:quench kinetic}
\end{figure*}

\subsection{Results of Chemical kinetic modelling} \label{sec:kineticsRes}

Temporal variation of \ce{H2O}, \ce{CO2}, and \ce{CH4} vertical abundances at the equator and longitude $\phi=225\degree$\revisionone{, as an example,} are shown in \hyperref[fig:gCV profiles]{Fig.}\;\ref{fig:gCV profiles} (a, d, g) to demonstrate how disequilibrium processes change the composition of the upper atmosphere (see \hyperref[sec:kineticmodelling]{Section}\;\ref{sec:kineticmodelling} for a description of the chemical kinetic model). The temporal evolution of the molecular abundances of \ce{H2O}, \revisionone{\ce{CO2}} and \ce{CH4} are color-coded by time where blue to red shows the progress in time from thermochemical equilibrium to their disequilibrium steady-states. All three molecular abundances depart from thermochemical equilibrium values at lower pressures and at some point remain constant up to the top of the atmosphere. \ce{CH4} quenches at around 0.1\;bar for this particular trajectory (\hyperref[fig:gCV profiles]{Fig.}\;\ref{fig:gCV profiles} (g)).

The quenching level of \ce{H2O}, however, can not be estimated easily and at the first glance could be somewhere between 0.3 and 30\;bar. Its thermochemical equilibrium abundance is almost constant through this region and hence diffusion does not significantly change its abundance below 0.3\;bar. \hyperref[fig:gCV profiles]{Fig.}\;\ref{fig:gCV profiles} (d) shows yet another complication with the quenching level determination. In this case, \ce{CO2} departs from its thermochemical equilibrium value at around 0.3\;bar, while it does not reach a constant value until around 3\;mbar. This usually occurs when the chemical and mixing timescales are comparable in that region. Examining abundance variation of all species at all trajectories reveals that in many cases the abundances do not reach a constant value at lower pressures and thus no quenching level (and consequently no ``quenched-abundance'') can be defined. Now, the question is, {\it what should be reported as the quenching levels?}

We follow \citet{molaverdikhani2019cold2} in suggesting the use of the `Coefficient of Variation (CV)' to estimate the deviation from chemical equilibrium. This quantity can be calculated as the ratio of the standard deviation of a given species' abundance, $s$, to its temporal mean abundance at any given pressure level.  CV is a relative and dimensionless measure of dispersion and hence independent of the absolute value of the mixing ratios. For each 1D trajectory we can therefore calculate a 1D CV profile. Since in our kinetic model the time step for the numerical integration is exponential, we use the geometric Coefficient of Variation (gCV) of abundances for this purpose, calculated as follows \citep{schiff2014head}:

\begin{equation}
{\displaystyle {\rm gCV}={\sqrt {\mathrm {e} ^{s_{\rm {ln}}^{2}}-1}}},
\label{eq:gCV}
\end{equation}
where ${s_{\rm ln}}$ is the scaled natural log of the standard deviation of abundances and can be estimated as $s_{\rm ln} = s \ln(10)$. \hyperref[fig:gCV profiles]{Fig.}\;\ref{fig:gCV profiles} (b, e, h) show the calculated gCV profiles of \ce{H2O}, \ce{CO2}, and \ce{CH4}, respectively, for latitude $\theta=0\degree$/ longitude $\phi=225\degree$ trajectory as an example. A higher gCV value represents a stronger variation of abundance at any given pressure level. A choice of gCV=0.05 seems to reasonably separate the chemical equilibrium and disequilibrium regions. The geometric mean of gCV for all species, $\widehat{\rm gCV}_{\rm all}$, can be also used to determine the quenching level of the atmosphere:

\begin{equation}
{\displaystyle \widehat{\rm gCV}_{\rm all} = \left(\prod _{i=1}^{n} \rm gCV_{i}\right)^{\frac {1}{n}}}.
\label{eq:gCVall}
\end{equation}

The gCV can be thought of as a measure of normalised abundance deviation from its mean value and within a given time period. Therefore, the chosen time interval during which the abundances are considered should start at a time at which the atmosphere is at the thermochemical equilibrium and end after reaching a new steady state, such as diffusion equilibrium.

Transection gCV-maps of \ce{H2O}, \ce{CH4}, and for all species are shown in \hyperref[fig:quench kinetic]{Fig.}\;\ref{fig:quench kinetic}. Interestingly, the zeroth-order approximation results (red solid lines in \hyperref[fig:quench kinetic]{Fig.}\;\ref{fig:quench kinetic} right-panels) are in good agreement with our kinetic modeling results, although it generally underestimates the depth of quenching on the dayside. An improved zeroth-order approximation can be foreseen as a fast method to estimate the quenching levels in parameterized atmospheric simulations.

The gCV values can be also used to compare the effectiveness of disequilibrium processes on different atmospheric constituents. For instance, \ce{H2O} has been driven away from thermochemical equilibrium less effectively than \ce{CH4} given gCV\textsubscript{\ce{H2O}}<gCV\textsubscript{\ce{CH4}}, \hyperref[fig:quench kinetic]{Fig.}\;\ref{fig:quench kinetic}; left and middle panels. $\widehat{\rm gCV}_{\rm all}$ values also hint that most of the species have been driven out of thermochemical equilibrium more effectively than these \revisionone{two} species on the night-side at the equatorial plane. \revisionone{\ce{CH4} equatorial map shows an asymmetry of gCV\textsubscript{\ce{CH4}} with a maximum value around longitude of 225$\degree$. This is due to lower temperatures on the nighside (Fig.~\ref{fig:qp0th} and Fig.~\ref{fig:quench0th maps}) and higher vertical mixing (Fig.~\ref{fig:quench0thKzzPq}) at this longitude at the equator.}

The noon-midnight gCV-maps (\hyperref[fig:quench kinetic]{Fig.}\;\ref{fig:quench kinetic} middle-panels) support shallower quenching levels on the dayside and a slight latitudinal dependency on the night-side. The anti-substellar point (Lat=0$\degree$, Lon=180$\degree$) appears to behave slightly different relative to its neighbouring trajectories, mostly due to stronger vertical velocity ($w$) at that region.

The deepest quenching levels occur at the polar regions, where the temperature is lower and the mixing is relatively strong. The $\widehat{\rm gCV}_{\rm all}$ map suggests a morning-evening asymmetry where the morning-side abundances have been driven more significantly out of thermochemical equilibrium and at deeper pressure levels. This asymmetry can potentially shape the transmission spectra of the two limbs differently. Observational consequences of chemical disequilibrium will be discussed in a forthcoming paper.

\subsection{The effect of photochemistry on the quenching levels} \label{sec:photochem}
Photochemistry is another disequilibrium process. The main source of photolysis on highly irradiated planets is the irradiation from the host star. Other sources (such as cosmic rays, GCR, stellar background UV radiation, or solar Lyman-$\alpha$ photons that are scattered from atomic hydrogen in the local interplanetary medium, LIPM) could be also important on weakly irradiated planets, and can be neglected in the case of HAT-P-7b. But how much does the photochemistry affect the quenching levels on a highly irradiated planet?

\begin{figure}
\includegraphics[width=\columnwidth]{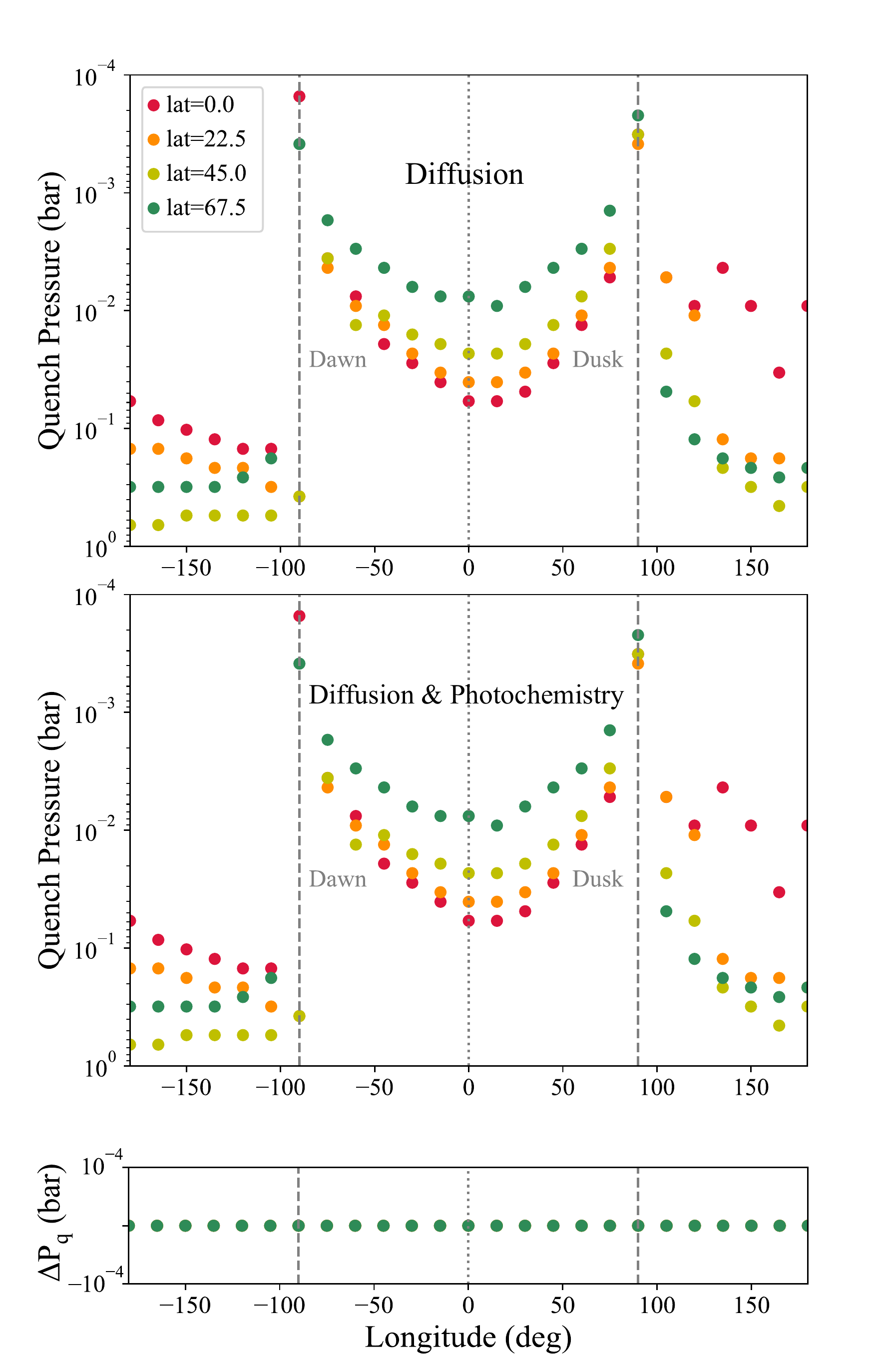}
\caption{{\bf Top)} Latitude and longitude dependency of quenching pressures, calculated from kinetic simulations by only including diffusion. The quenching criterion is discussed in  \hyperref[sec:kineticsRes]{Section}\;\ref{sec:kineticsRes} {\bf Middle)} Similar to the top panel but photochemistry is also considered. {\bf Bottom)} The difference between the quenching pressures of models with only diffusion and models including both diffusion and photochemistry. No difference is noticeable; suggesting photochemistry does not play a significant role in the quenching levels of HAT-P-7b.}
\label{fig:quench photo}
\end{figure}

\hyperref[fig:quench photo]{Figure}\;\ref{fig:quench photo} compares the latitudinal quenching levels, $\widehat{\rm gCV}_{\rm all}$, with and without considering the photochemistry. The bottom panel shows the differences and suggests that the photochemistry plays a negligible role. This is not surprising since the photochemistry mostly affects the mbar regime and higher altitudes (see e.g. \citet{molaverdikhani2019cold2} and references therein) and HAT-P-7b's dayside quenching points are located at deeper levels. The two exceptions are the evening terminator and a portion of the morning terminator where the quenching pressures can be lower than mbar, as discussed in Section~\ref{sec:zeroth_appx_res}. However, at these regions the zenith angle is 90$\degree$; hence irradiation is ignored in our 1D kinetic calculations. The nightside also does not change since GCR and LIPM are not included.

\subsection{The effect of mixing strength on the quenching levels} \label{sec:efficiency}

\begin{figure}
\includegraphics[width=\columnwidth]{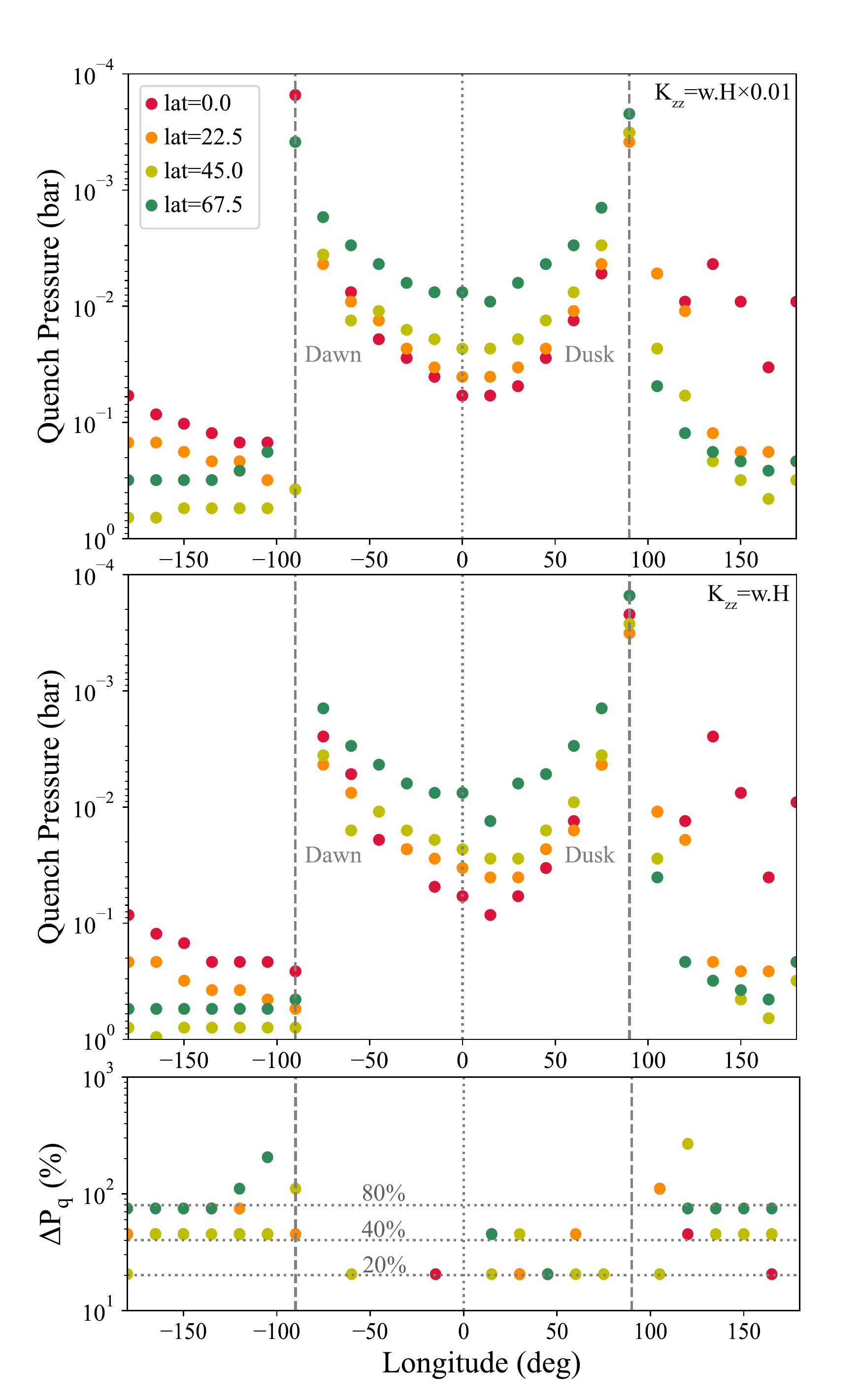}
\caption{{\bf Top)} Latitude and longitude dependency of quenching pressures, calculated using $K_{zz}$=w$\cdot H\times$ $10^{-2}$. {\bf Middle)} Similar to the top panel but using $K_{zz}=w\cdot H$, i.e. traditional approach. {\bf Bottom)} \revisionone{Comparing the quenching pressures of models with $K_{zz}=w\cdot H$ to the quenching pressures of models with $K_{zz}$=w$\cdot H\times$ $10^{-2}$.} The atmosphere is being quenched at deeper levels, usually 20\%-40\% on the dayside and 40\%-80\% on the nightside. Some terminator trajectories, however, show several orders of magnitude change in their quenching levels, deepening from the mbar to 1 bar pressures.}
\label{fig:quench kzz compare}
\end{figure}

As discussed in Section~\ref{sec:zeroth_appx}, we decreased K$_{zz}$ values by two orders of magnitude, Eq.~\ref{eq:eq_Kzz}, to follow \citet{parmentier20133d} conclusion. Using the traditional mixing strength approximation, i.e. without decreasing $K_{zz}$ by a factor of 0.01, would then result in different quenching levels. \hyperref[fig:quench kzz compare]{Figure}\;\ref{fig:quench kzz compare} compares the two cases and as one might expect the difference is non-negligible.

The resulted quenching levels from the traditional approach, $K_{zz}$=w$\cdot H$, are usually deeper than the modified one, $K_{zz}$=w$\cdot H\times$ $10^{-2}$. The differences are on the order of 20\%-40\% on the dayside and 40\%-80\% on the nightside. The quenching levels at the terminator trajectories, however, show stronger variations as high as several orders of magnitude.

\begin{figure*}
    \includegraphics[width=0.45\textwidth]{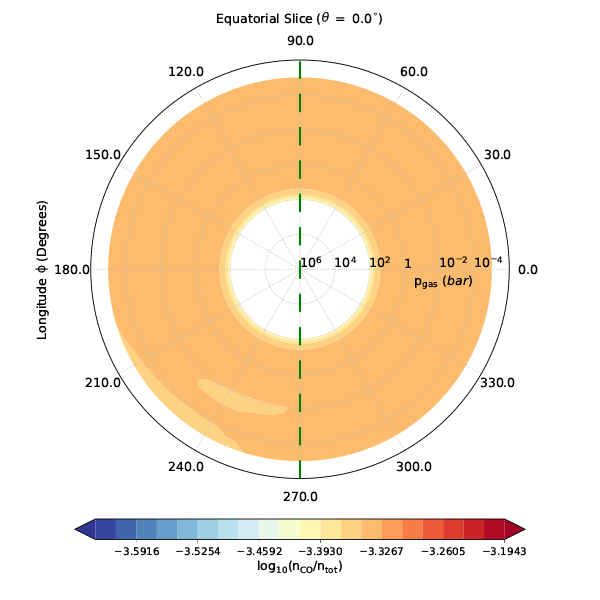}
        \includegraphics[width=0.45\textwidth]{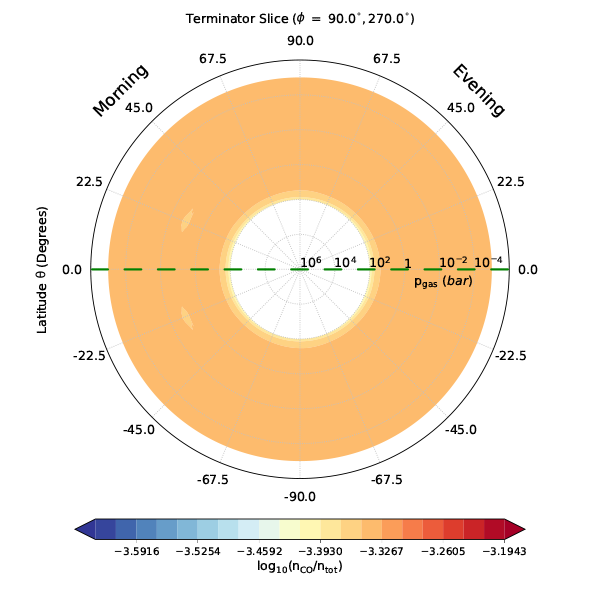}\\*[-0.4cm]
    \includegraphics[width=0.45\textwidth]{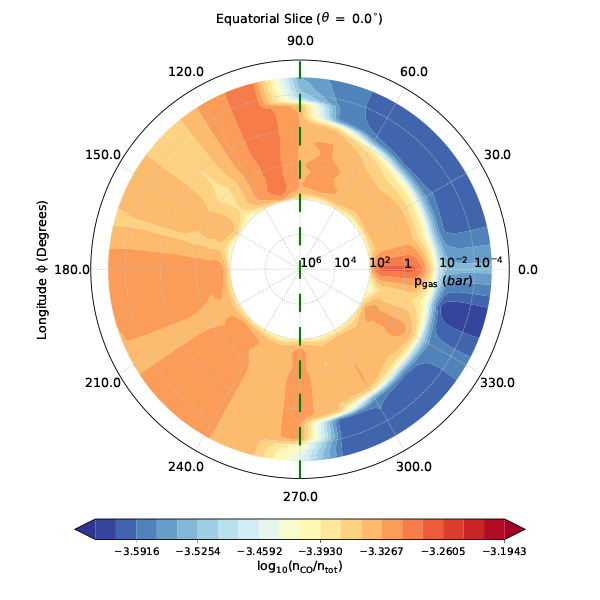}
        \includegraphics[width=0.45\textwidth]{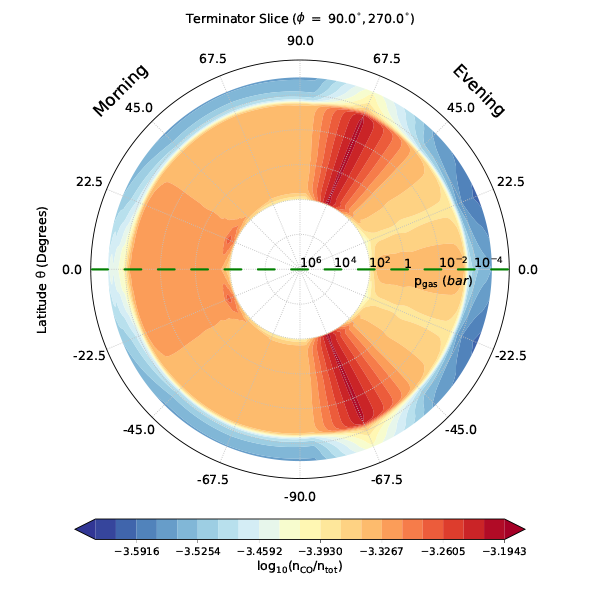}\\*[-0.4cm]
    \includegraphics[width=0.45\textwidth]{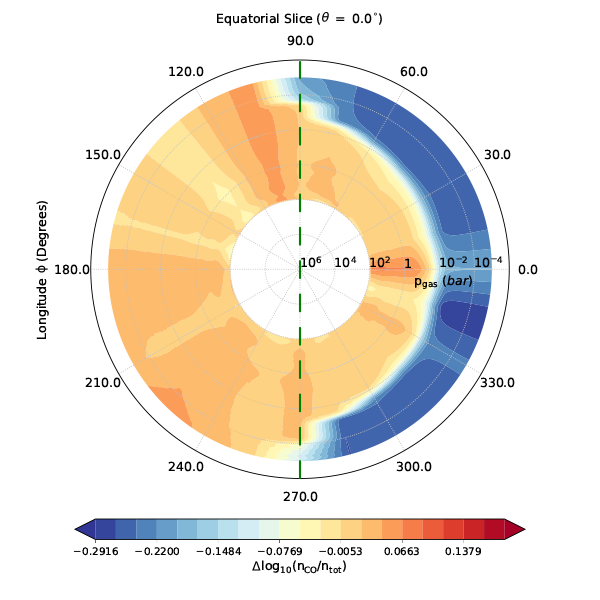}
       \includegraphics[width=0.45\textwidth]{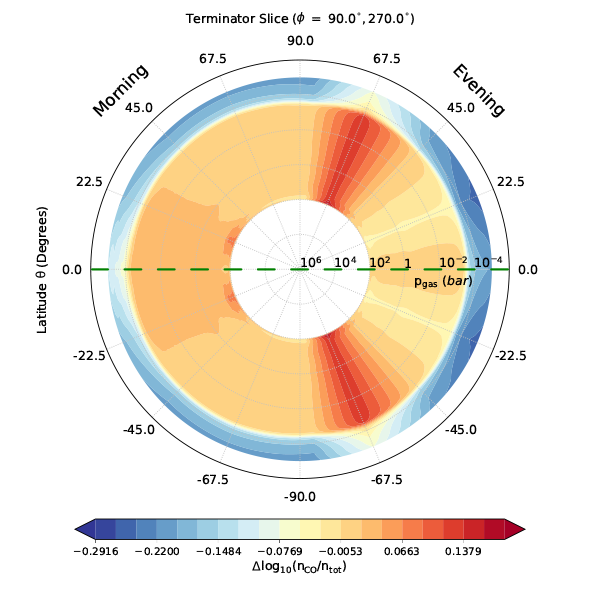}\\*[-1cm]
    \caption{Carbon monoxide (CO):  {\bf Left column:} equatorial slices ($\theta=0\degree$), {\bf Right column:} terminator slices; 
    \textbf{Top:} initial condition, $\log(n_{\rm CO}^{\rm init}/n_{\rm tot})$; \textbf{Middle:}  final result from kinetic simulation, $\log(n_{\rm CO}^{\rm final}/n_{\rm tot})$ \textbf{Bottom:} difference, $\log(n_{\rm CO}^{\rm init}/n_{\rm tot} - n_{\rm CO}^{\rm final}/n_{\rm tot}$ between the initial (equilibrium) values and the kinetic results.}
    %
    \label{fig:CO_results}
\end{figure*}

\section{Disequilibrium abundances due to Diffusion and Photochemistry} \label{s:maps}

\subsection{CO and \ce{CH4} abundance at equatorial and terminator regions}
In Fig. \ref{fig:CO_results} and \ref{fig:CH4_results} \revisionone{(in the Appendix)}, we present the differences of the CO and \ce{CH4} abundance of the kinetic model and the equilibrium condition in the equatorial (left) and terminator (right) regions. The initial conditions (top row) is set to the equilibrium results from Paper I where the element depletion due to cloud formation affects the gas-phase abundances in the cloud-forming regions of the atmosphere.

\revisionone{The carbon monoxide abundance  (Fig. \ref{fig:CO_results}) on HAT-P-7b is only marginally affected by quenching, and the kinetic modelling results vary by less than a factor of two compared to the number densities in chemical equilibrium for both equatorial and terminator regions.} On the other hand, the methane abundance (Fig.~\ref{fig:CH4_results}) in the kinetic model shows a stronger deviation from the equilibrium results at $\sim$10\;mbar level. The strong vertical mixing causes \revisionone{the number density of methane at P$<$1\;bar around longitude $\phi=240^{\circ}$ to decrease by a factor of about 1000. An enhancement in methane number density at $p_{\rm gas}>$1\;bar by more than 10 times is also noticeable.} At the morning terminator, there is a strong local methane enhancement in between 1-10\;mbar level. We interpret that this methane enhancement is related to the thermal inversion at $\sim$10\;mbar level (see Fig. \ref{fig:qp0th} and Fig. \ref{fig:quench0th maps} in the appendix). The vertical mixing homogenizes the relatively methane-rich layer of the cooler region at $p_{\rm gas} >$10\;mbar with the hotter region at $p_{\rm gas}<$10\;mbar.
As a result, the vertical mixing leads to a local enhancement in methane's number density between 1-10\;mbar level relative to that of equilibrium result.

\subsection{\ce{H2O} abundance at equatorial and terminator regions}

Similarly we show the effects on \ce{H2O} abundance (Fig \ref{fig:H2O_results}, as with carbon monoxide the abundance is relatively unchanged, continuing to display an asymmetry between the dayside and nightside. However, some subtle variations are notable. While abundances at pressures larger than 1\;bar mostly stay constant at all longitudes around the equator, abundances at higher altitudes on the nightside increase slightly. On the dayside, the abundance difference between equilibrium and kinetic chemistry is almost invariant at pressures larger than 100\;mbar and at higher altitudes it decreases by \revisionone{a factor of about 30} on average.

The terminator also shows only small variations. Albeit a prominent change being the shifting of relatively low abundance regions from around 0.01\;bar between $\phi\pm$ 67.5\;degrees latitude at the morning side to two vertical columns at the same latitudes on the evening side. The most significant change of a factor of nearly three orders of magnitude decrease occurs just on the dayside of the morning terminator ($\theta\sim 285$ degrees longitude) at 1-10\;mbar, this is likely linked to the thermal inversion there as was the case for methane.

We should note that the abundance variation results due to disequilibrium are consistent with the interpretations based on gCV calculations, as discussed in Section~\ref{sec:kineticsRes} for the two cases of \ce{H2O} and \ce{CH4}. Therefore, gCV can be used both for the determination of quenching pressures and as a quantitative measure of the efficiency of disequilibrium processes on the atmospheric constituents.

\subsection{CH, \ce{C2H2}, CN,  HCN abundance at equatorial and terminator regions}

Here we show how the abundances of small carbon molecules CH (methylidyne, Fig.~\ref{fig:CH_results}) and  CN (cyanide, Fig.~\ref{fig:CN_results}) may be affected by vertical transport processes compared to the chemical equilibrium results in Paper I. We note, however, that the atmosphere is globally oxygen-rich (bulk carbon-to-oxygen ratio (C/O)=0.54), hence, oxygen-binding molecules will dominate the atmosphere's chemical composition. It is, however, worthwhile to consider carbon-molecules that can be of relevance for the formation of biomolecules (like \ce{CH4}, \ce{C2H2} and HCN) and their precursors in order to identify possible tendencies \revisionone{in} the unexplored parameter ranges that exoplanet provide. HCN is furthermore suggested as spectral tracer for lightning events in planetary atmospheres. Knowing its background abundance may help to disentangle such lightning traces \citep{hodosan2017exo,helling2019lightning}.

CN and CH are precursor for more complex molecules as well; such as \ce{C2H2} (Fig.~\ref{fig:C2H2_results}) and HCN (Fig.~\ref{fig:HCN_results}).  CH, CN, \ce{C2H2} and HCN are molecules which typically appear in spectra of carbon-rich stars \citep{eriksson1984effects,helling1996formation} and which therefore maybe considered as potential tracers for carbon-enriched planetary environments (e.g., \citealt{1998A&A...335L..69H,kumar2011photochemical,madhusudhan2012c,moses2013compositional,venot2015new,2016A&A...585A.145U,2016ApJ...828...15H}).

Firstly, CH and CN show a clear day/night asymmetry with both being more abundant on the dayside and in the warmer atmospheric regions. This  day/night asymmetry is unaffected by quenching as it is driven by the local day/night temperature difference instead of hydrodynamic motions. The CH abundance has decreased on the nightside compared to the thermochemical equilibrium results. Differences are generally small but amount to orders of magnitudes in localised areas in the equatorial plane as much as \revisionone{a factor of $10^{12}$ (i.e. 12\;dex)}. Differences remain negligible in the low-pressure terminator regions ($p_{\rm gas}<1$bar).

\ce{C2H2} is very low abundant and kinetic chemistry only affects its abundances in localised areas in the terminator regions at $p_{\rm gas}\approx 10^{-2}$bar. \ce{HCN} is somewhat more abundant than \ce{C2H2}, but disequilibrium effects introduce a day/night asymmetry for HCN. As a result, the nightside \ce{HCN} abundance has increased by 3 orders of magnitude compared to the chemical equilibrium results. The absolute number densities, however, remain small compared to CO and \ce{H2O}, for example. This asymmetry is also visible in the terminator regions where the morning terminators show a higher HCN abundance. The terminator regions (right column) is again only locally affected by the kinetic gas chemistry, mainly at the nightside at $p_{\rm gas}\approx 10^{-2}$\;bar. In general, we note that all molecules in the form of C$_n$H$_m$ are affected by disequilibrium processes on the nightside, where heavier C$_n$H$_m$ molecules are affected the most.

The result in this section emphasis our finding from the previous sections: Quenching of the gas phase chemistry does take place for some of the molecules but the actual deviation needs to be considered with care as large deviations can be local phenomena only.

\begin{figure*}
\includegraphics[width=0.45\textwidth]{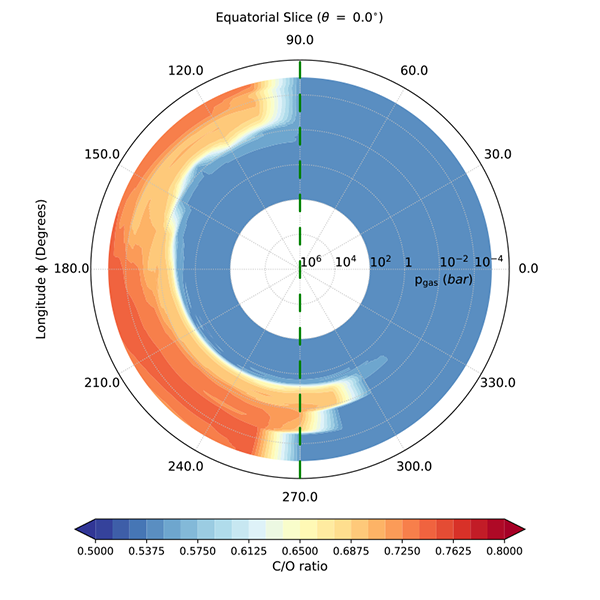}
    \includegraphics[width=0.45\textwidth]{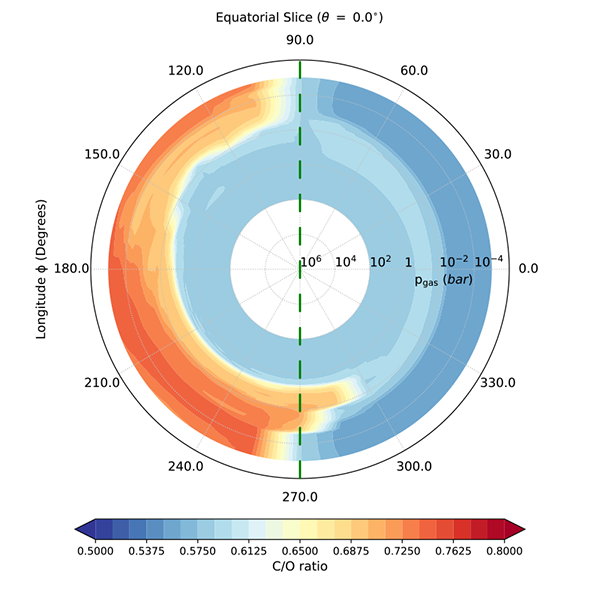}\\
    \includegraphics[,width=0.45\textwidth]{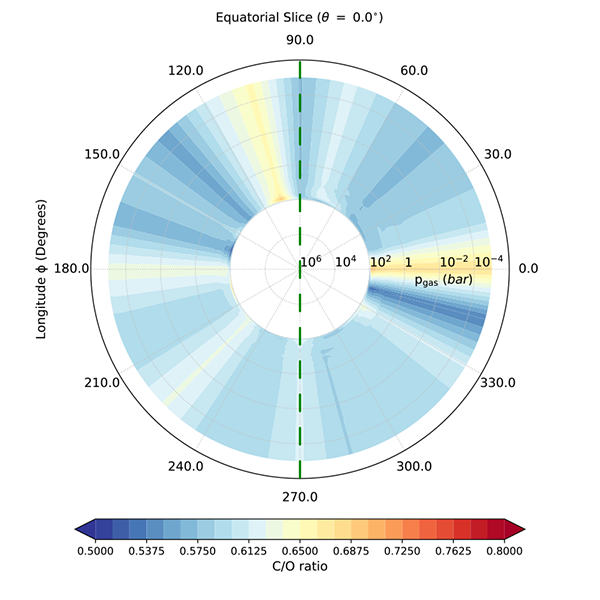}
    \includegraphics[width=0.45\textwidth]{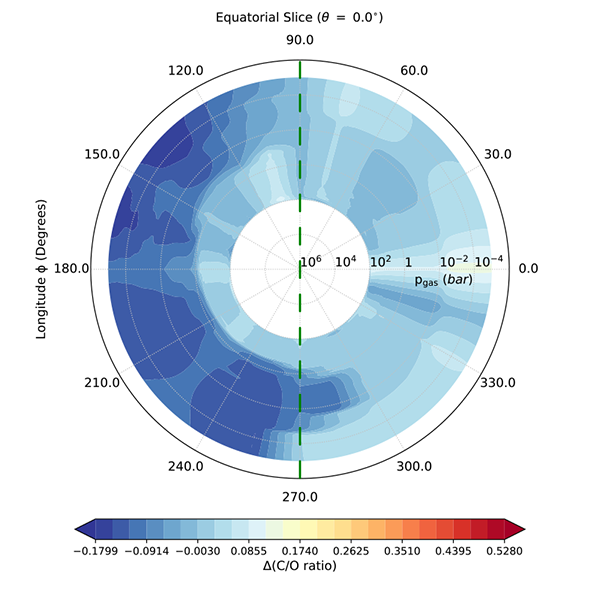}
    \caption{The carbon-to-oxygen ratio (C/O): \textbf{top left:} C/O after cloud formation as in Paper I according to $\epsilon_{C}/\epsilon_{O}$, 
    \textbf{top right:} C/O for initial condition of gas-kinetic calculations (from CH, CO, \ce{CO2}, \ce{C2H2, ...), },  
    \textbf{bottom left:} C/O at the end of gas-kinetic calculations (t=10$^{12}$\;sec) \textbf{bottom right:} C/O change between initial condition (top right) and final gas-kinetic solution (bottom left).}
        \label{fig:CO_equ_Ratio_TEST}
\end{figure*}

\subsection{The atmospheric C/O in HAT-P-7b}\label{ss:CtoO}

\revisionone{The carbon-to-oxygen ratio (C/O) from $\ll$ 1 to $\gg$1 has served as one of the stellar parameters which allowed the link to stellar evolution and the spectral changes have widely been studied (\citealt{2008A&A...486..951G,2017A&A...601A..10V}). Its effects on the spectra of planetary atmosphere has been also studied extensively \citep[e.g.][]{seager2005dayside,madhusudhan2012c} and it has shown to change the atmospheric properties significantly \citep[e.g.][]{molliere2015model,molaverdikhani2019cold,molaverdikhani2019cold2}. Thus,} C/O ratio became one of the parameters that was hoped to be useful to characterise an exoplanet's formation and/or evolution.

\revisionone{However,} C/O can not be straight forwardly linked to planetary evolutionary states nor to planet formation easily if clouds form \citep{2014Life....4..142H}. We note that other mineral ratios (like Mg/O, Mg/Si etc.) will be similarly unsuitable as a direct link to planet formation unless cloud formation and related chemical processes are simulated simultaneously (see e.g. Fig. 4 in \citealt{doi:10.1146/annurev-earth-053018-060401}).

The main contribution to changes of C/O in our simulations comes from the change in oxygen abundance due to consumption in cloud particles. Our cloud formation simulations for HAT-P-7b start from solar element abundances with C/O=0.54, i.e. more oxygen than carbon element abundance.  The resulting C/O for the cloud-forming atmosphere of HAT-P-7b linked to the equilibrium gas-phase code \revisionone{GGChem \citep{woitke2018equilibrium}\footnote{GGChem code is publicly available at https://github.com/pw31/GGchem}} is reproduced in Fig.~\ref{fig:CO_equ_Ratio_TEST} (top left) for the equatorial slice ($\theta=0\degree$): The cloud forming nightside reaches C/O$\approx 0.78$ and the cloud free parts of the dayside remain at the undepleted, solar value as expected. The oxygen and carbon element abundances that result from our kinetic gas simulations are shown in Fig.~\ref{fig:CO_equ_Ratio_TEST} (bottom left).

We compare in how far the undepleted, solar C/O ratio is reflected in the initial conditions for the gas-kinetic analysis conducted in this paper. Then, we study how C/O may have been affected by chemical gas-kinetic processes. We present our results in the form of slice plots through the equatorial plane in Fig.~\ref{fig:CO_equ_Ratio_TEST}. The corresponding terminator slices can be found in the appendix (Fig.~\ref{fig:CO_equ_Ratio_terminator}).

The C/O ratio used as initial condition for the kinetic modelling (top right in Fig.~\ref{fig:CO_equ_Ratio_TEST}) is smaller than the C/O derived in Paper I (top left in Fig.~\ref{fig:CO_equ_Ratio_TEST}). The reason is that such rate networks consider less molecules than, e.g., our gas-phase equilibrium calculations with GGChem, and hence, C/O is derived from only the C and O binding molecules considered in the gas-kinetic network simulation (see list in Sect.~\ref{sec:kineticmodelling}). The input abundances for these molecules were set to those resulting from our cloud formation calculations in Paper I, as discussed in Set. ~\ref{sec:kineticmodelling}.

The gas kinetic C/O ratios appear to be relatively homogenised toward a value slightly higher than its solar value. This is expected as the global initial C/O ratio in the gas phase was higher than the solar value due to the depletion of oxygen by cloud formation. The largest differences between (C/O)$_{\rm eq}$ and (C/O)$_{\rm kin}$ appear for p$_{\rm gas}<$1\;bar on the nightside and to some extend on the morning terminator. All other regions show negligible deviations between (C/O)$_{\rm eq}$ and (C/O)$_{\rm kin}$ in the equatorial plane. This homogenisation of C/O ratio removes the morning-evening C/O asymmetry due to the cloud formation only (reported in Paper\;I). This suggests a C/O asymmetry may be less pronounced in the ultra-hot jupiters, however, an observed global enhancement of C/O ratio could be still an indication of cloud formation on a hot exoplanet.

\subsection{Observability of disequilibrium chemistry in HAT-P-7b's atmosphere}\label{ss:gas_opa}

Having examined how molecular abundances are modified by accounting for chemical kinetics, we now turn to the prospect of observing the influence of disequilibrium chemistry \revisionone{on the atmospheric spectra}. There are many ways in which disequilibrium chemistry can manifest in an exoplanet spectrum. The clearest diagnostic would be the appearance of absorption features caused by a molecule expected to be undetectable for equilibrium chemical abundances. For example, signatures of NH$_3$ and HCN can become detectable in hot Jupiter transmission spectra if disequilibrium mechanisms enhance their abundances to $\gtrsim 1 \%$ that of H$_2$O \citep{MacDonald2017b}. However, such a dramatic change requires molecular abundances to alter at the order-of-magnitude level. Alternatively, for smaller deviations from chemical equilibrium, one may examine whether the shape of a molecule's absorption features, or the relative strengths of features due to different molecules, are altered by chemical kinetics. 

We quantitatively asses the influence of disequilibrium chemistry on spectral observations of HAT-P-7b using a similar approach to Paper I. Specifically, we compute the pressure at which the vertically integrated optical depth due to a chemical species, $x$, reaches unity (i.e. $p (\tau_x (\lambda)=1)$). This provides a useful indication of the contributions of different molecules to spectra of HAT-P-7b. We first consider which of the molecules included in our chemical model are likely to be spectrally accessible to observations. We then examine differences due to chemical kinetics, quantified by the \% difference in $p (\tau_x (\lambda)=1)$ between the initial and final atmospheric abundances.

\begin{figure*}[ht!]
    \centering
    \includegraphics[width=\textwidth]{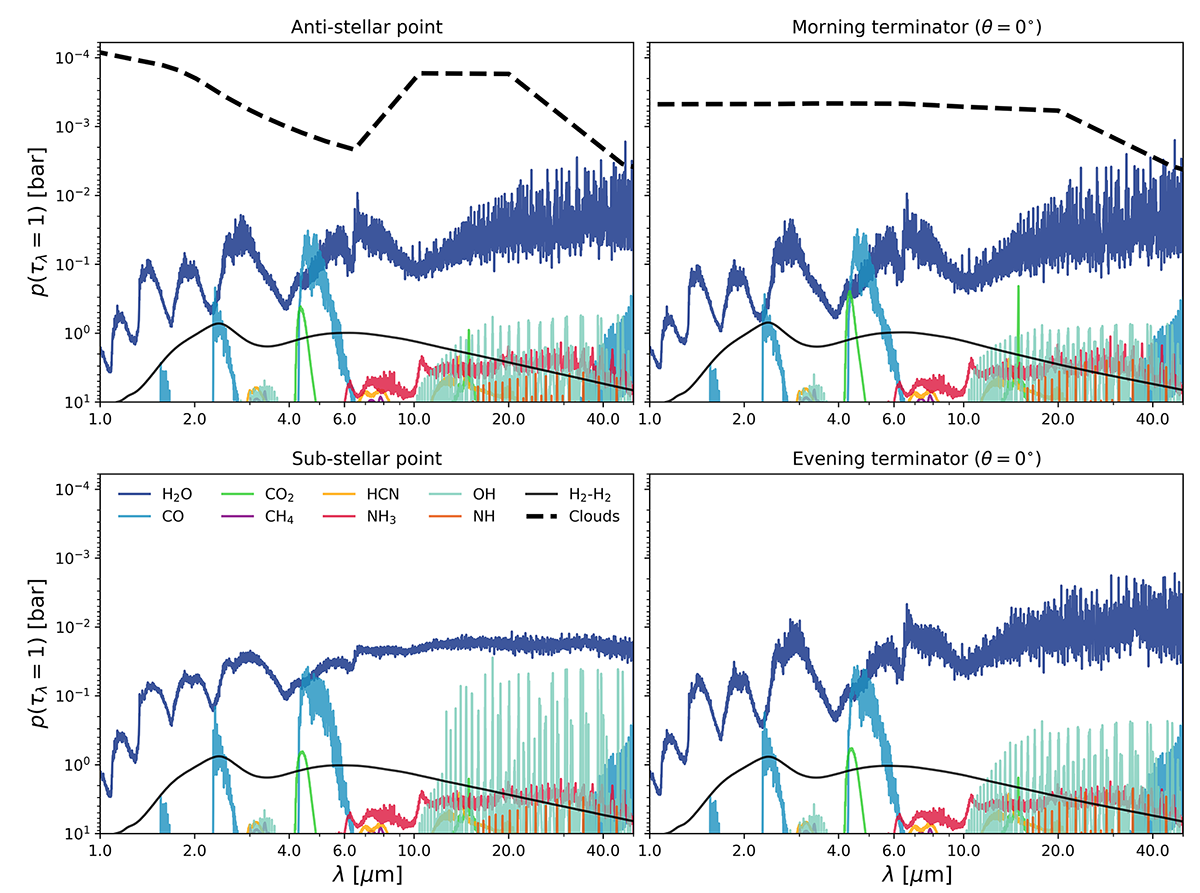}
    \caption{Prominent sources of molecular opacity under disequilibrium chemistry. The atmospheric pressure at which the vertical wavelength-dependent optical depth of species $x$ reaches unity, $p (\tau_x (\lambda)=1)$, is shown across the infrared. Several molecules included in both the kinetic model and the opacity database which do not reach $\tau_x (\lambda) = 1$ by the 10\,bar level (e.g. C$_2$H$_2$) are omitted. See Paper I (Fig. 20) for opacity contributions due to species not included in the kinetic model (e.g. heavy-metal oxides, atoms, etc.). All opacities are plotted at a spectral resolution of $R = 1000$. Four equatorial regions are considered: (i) the anti-stellar point, (ii) the morning terminator, (iii) the sub-stellar point, and (iv) the evening terminator. For comparison, the $\tau (\lambda) = 1$ surface due to clouds from Paper I is overlaid.}
    \label{fig:P_tau_Kinetic}
\end{figure*}

Figure~\ref{fig:P_tau_Kinetic} shows the $p (\tau_x (\lambda)=1)$ surfaces for prominent molecules at the end of the kinetic calculation. Here we focus on infrared wavelengths, as the opacity at visible wavelengths will contain large contributions from atomic and heavy-element molecules not included in the present kinetic chemistry model (see Paper I, Fig. 20, for their influence in chemical equilibrium). Three molecules are potentially observable at near-infrared wavelengths: \ce{H2O}, CO, and \ce{CO2} (ordered by prominence). The importance of these molecules holds in all four of the plotted atmospheric regions, though their relative influence does vary (e.g. CO and CO$_2$ opacity is more prominent, compared to H$_2$O, at the morning terminator than the evening). At the sub-stellar point, OH also becomes an important opacity source for wavelengths longer than $> 10~\mu$m (opening the possibility of its detection via mid-infrared dayside emission spectroscopy, such as with JWST's MIRI MRS mode -- see \citet{Beichman2014}). While $H^-$ is reported to be an important opacity source on the highly irradiated exoplanets 
\citep[see e.g.][]{arcangeli2018h}, we found that its contribution \revisionone{(at thermochemical equilibrium)} is likely to be significant only on the dayside of HAT-P-7b and at the wavelengths longer than 15\;$\mu$m \citep{paperI}; suggesting no significant impact of this compound on the transmission spectra of this planet.

It is important to note that molecular opacity at the anti-stellar point and morning terminator are likely to be obscured by clouds (see Paper I). This overall picture is essentially the same as Paper I, indicating that relaxing chemical equilibrium does not alter the prediction for which molecules are expected to be most prominent in spectra of HAT-P-7b.

\begin{figure*}[ht!]
    \centering
    \includegraphics[width=\textwidth]{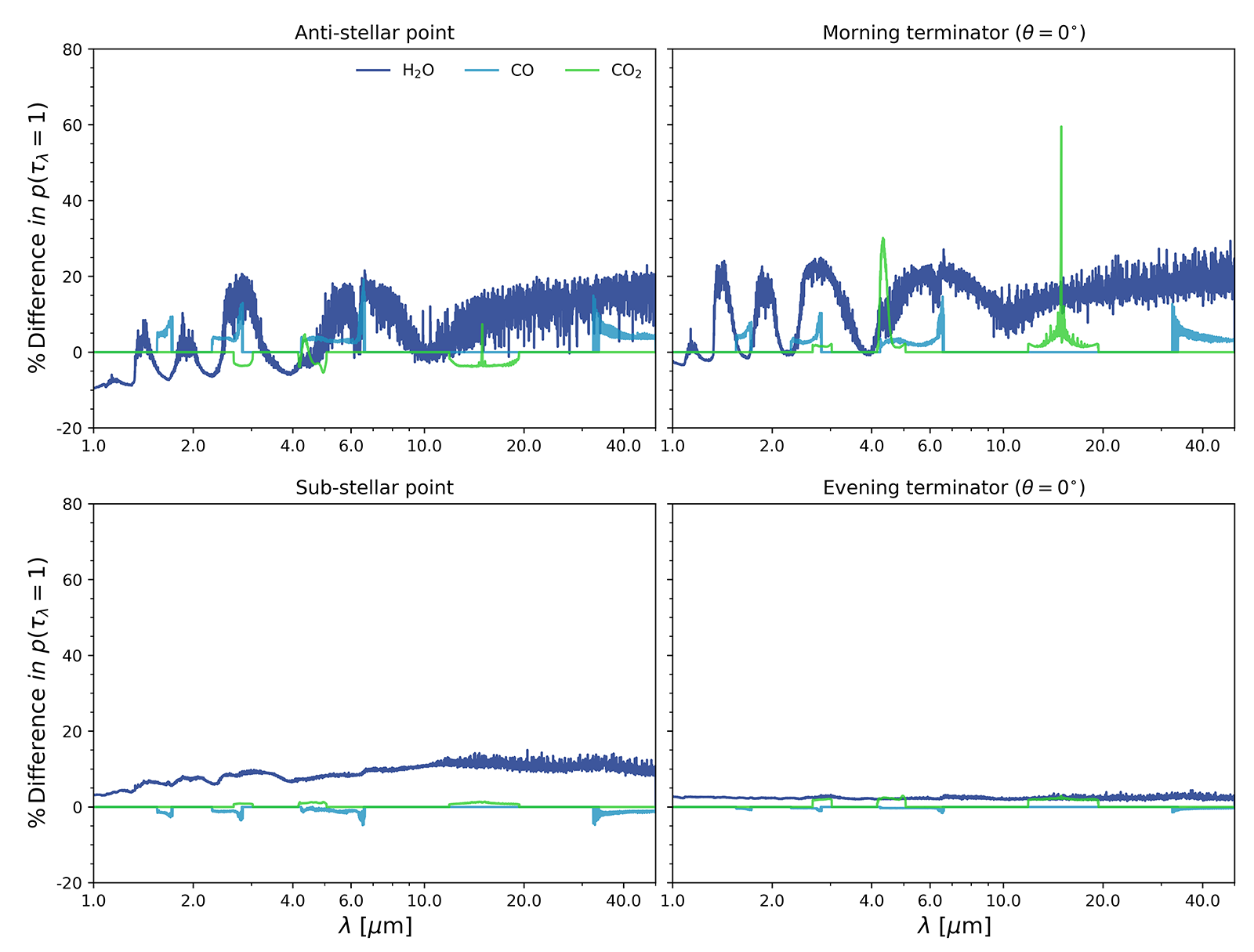}
    \caption{Differences to the opacity contribution of observable molecular species due to disequilibrium chemical kinetics. The difference metric, $(p (\tau_x (\lambda)=1)_{\rm Initial} - p (\tau_x (\lambda)=1)_{\rm Kinetic}) / p (\tau_x (\lambda)=1)_{\rm Initial}$, is defined such that a positive difference implies stronger absorption after accounting for disequilibrium chemistry (i.e. $p (\tau_x (\lambda)=1)$ occurs higher in the atmosphere). Only H$_2$O, CO, and CO$_2$ are plotted, as these molecules offer the greatest detection prospect in the near-infrared (see Figure~\ref{fig:P_tau_Kinetic}). All opacities are plotted at a spectral resolution of $R = 1000$. Four equatorial regions are considered: (i) the anti-stellar point, (ii) the morning terminator, (iii) the sub-stellar point, and (iv) the evening terminator.}
    \label{fig:P_tau_Difference}
\end{figure*}

However, as we have seen, chemical kinetics plays an important role in shaping the vertical abundance profiles of various molecules. For example, Figure~\ref{fig:gCV profiles} demonstrates nightside quenching of the H$_2$O abundance in the upper atmosphere for pressures $\lesssim 10^{-1}$\,bar. Given that H$_2$O is the most prominent molecular opacity source at infrared wavelengths, any such abundance changes at atmospheric pressures accessible to transmission and emission spectra are potentially observable. We examine the importance of disequilibrium abundance changes by computing the \% difference between the initial (equilibrium) and final (kinetic) $p (\tau_x (\lambda)=1)$ surfaces for the molecules anticipated to be most prominent in spectra of HAT-P-7b. The results are shown in Figure~\ref{fig:P_tau_Difference}.

Disequilibrium chemistry tends to manifest by enhancing the prominence of molecular absorption features. The differences are most pronounced at the anti-stellar point and morning terminator, where the $p (\tau_x (\lambda)=1)$ surfaces occur up to 20\% higher in the atmosphere for H$_2$O (and up to 60\% higher for CO$_2$) in the centre of absorption features (e.g. $3~\mu$m) and up to 10\% lower in absorption minima (e.g. $1.7~\mu$m). This disequilibrium-induced contrast in the pressures where the atmosphere becomes optically thick between wavelengths where molecules strongly and weakly absorb would result in `stretched' spectral absorption features. On the other hand, at the sub-stellar point and evening terminator the $p (\tau_x (\lambda)=1)$ surface for H$_2$O increases in altitude for all wavelengths, which would result in stronger, but relatively unstretched, absorption features. We thus see that, in principle, disequilibrium chemistry can transform the shape of molecular absorption features in different ways for different atmospheric regions. Although the anti-stellar point tends to be inaccessible to observations (due to its low thermal emission), these differing vertical opacity distributions between the morning terminator (observed in transmission) and the sub-stellar point (observed in emission) could be one avenue to probe disequilibrium chemistry in ultra-hot Jupiters. A second avenue is inter-terminator differences observed in transmission spectra, due to the morning terminator experiencing enhancements in H$_2$O, CO, and CO$_2$ absorption whilst the evening terminator sees only minor changes. Ultimately, whether or not these disequilibrium-induced opacity differences will be observable depends crucially on the presence of clouds in HAT-P-7b's atmosphere. If the morning terminator is dominated by cloud opacity (as predicted in Paper I), it will be difficult to observationally probe disequilibrium chemistry in this region. Dayside emission spectra may have easier time detecting H$_2$O features due to its enhanced abundance, but observing this alone would be difficult to uniquely attribute to disequilibrium chemistry (as opposed to, say, a higher atmospheric O/H ratio but in chemical equilibrium). We therefore conclude that observationally detecting signatures of disequilibrium chemistry in HAT-P-7b's atmospheres will be challenging. A quantitative assessment of the detectability of disequilibrium signature on HAT-P-7b will be present in \revisionone{a forthcoming paper}.


\section{Discussion} \label{s:discussion}
\revisionone{Here we discuss the limitations of our work and the caveats that must be taken into account when interpreting the results.}

\paragraph{Material properties: rate and diffusion coefficients.} Gas-kinetic rate networks solve diffusion equations, one for each molecular species (e.g. Eq. 22 in \cite{2016ApJS..224....9R}). Each equation requires material constants in the form of a rate, an eddy diffusion coefficient, and a molecular diffusion coefficient. In principle, they all can be measured through lab experiments, except, measuring them for many reactions/reactants is an overwhelming task. We assume that the rate values are reasonably well known (or have been benchmarked) and do not investigate the propagation of uncertainties in the rate coefficients into our results. The eddy diffusion is assumed to be the same for all species. The molecular diffusion of species are calculated following the Lennard-Jones Theory (see Appendix \ref{s:appendix_mol_diff} for a detailed description). This is a critical point for larger molecules in low-pressure regions where the molecular diffusion becomes more efficient than the turbulent eddy diffusion.

\paragraph{Network size:} Gas-kinetic simulations are limited by the number of reactions included. The more focused a network is, i.e. reduced network, the faster it will be. The ultimate form of such reduced network manifests itself in the chemical timescale relaxation scheme where only one rate limiting reaction is considered. The focus of such simulation is usually on a chemical inversion between two chemical compounds only. This is the most frequently used technique to couple 3D GCM simulations with chemical disequilibrium processes.

As shown by, e.g., \citet{drummond2018b} a consistent coupling of 
dynamics, radiative transfer, and chemistry demand more computational power and may result in uncovered physics by the relaxation scheme. For instance, an outcome of such full coupling leads to the wind-driven chemistry that in turn has a significant impact on thermal and dynamical structure of the cool atmospheres, such as HD\;189773b. The effect of full network coupling, however, reported to be less significant in the case hotter atmospheres such as HD\;209458b. Is this also the case for the ultra-hot Jupiters or not is a matter of question that requires a full coupling of chemistry and GCM.


\paragraph{Limitations of 1D kinetics models:} One important limitation of our approach is that our chemical kinetic model is one-dimensional. Thus, horizontal mixing is neglected. \revisionone{Horizontal mixing can homogenize abundances longitudinally \citep[e.g.][]{agundez2014} and/or latitudinally and lead to horizontal quenching. Recent studies coupling a simplified chemical relaxation scheme to 3D general circulation models of hot Jupiters have found that horizontal mixing dominates over vertical mixing \citep{mendonca2018b}. Coupling 3D GCM with reduced chemical networks also confirms this and showed that abundances are determined by a combination of vertical and horizontal mixing \citep{drummond2018b,2018ApJ...869...28D}.}

However, all of these studies focused on hot Jupiters with equilibrium temperatures $< 1500$\;K and it is not clear to what extend those results apply to ultra-hot Jupiters. \citet{Komacek2017} demonstrate that with increasing equilibrium temperature, day- to night flow becomes more important compared to the equatorial jet and vertical velocities increase. As a consequence, the efficiency of vertical mixing is believed to increase by more than an order of magnitude between $T_{eq}=$1500\;K and $T_{eq}=$2200\;K \citep{Komacek2019verticalmixing}. On the other hand, vertical velocities are set by the horizontal ones through the mass conservation. Therefore, in principle, stronger vertical velocities should result in stronger horizontal ones. Future studies applying GCMs including simplified chemical schemes, \revisionone{such as a new reduced network recently proposed by \citet{venot2019reduced},} to a wider range of planets will help clarify this aspect. In the long term, coupling full chemical networks to GCMs will hopefully become possible.

A related, but different limitation is that vertical mixing in kinetic models is parametrized as an eddy diffusion coefficient. \citet{ZhangShowman2018a,ZhangShowman2018b} demonstrate that this approximation is not always valid---the interaction of chemical species with the three-dimensional circulation can produce situations in which they do not behave diffusively. Furthermore, they find that species with different chemical lifetimes can have different eddy diffusivities, an effect that commonly has not been taken into account in the kinetic models.

We therefore stress that even though we use a state-of-the-art modelling approach, our results are meant to provide a first exploration of the extent to which disequilibrium chemistry plays a role on HAT-P-7b. Our results provide an estimate of the importance of disequilibrium abundances and in which regions of the atmosphere deviations from equilibrium chemistry can be expected. The derived disequilibrium abundances should, however, not be taken as exact.

\revisionone{\paragraph{Photodissociation data:} Simulation of photodissociation processes on warm/hot exoplanets require inputs relevant to these temperatures. However, measurements of such data, including cross-sections and branching ratios, are mostly available at around 300\;K. While neglecting temperature dependent cross-sections could change the compositions at the upper atmosphere of exoplanets profoundly \citep{venot2018vuv}, we showed that on the dayside of UHJs, photodissociation plays a negligible role. Due to relatively large uncertainties in the data used to model photodissociation, small differences in the stellar spectra (whether synthetic or observed) have also insignificant influence on the atmospheric composition.}

\paragraph{Photochemical hazes:} While the formation of photochemical hazes is poorly understood, it is typically assumed \citep[e.g.][]{rimmer2013cosmic,gao2017constraints,horst2018haze,edgington2018photochemistry,he2018photochemical,west2018temperature,kawashima2018theoretical,doi:10.1146/annurev-earth-053018-060401,berry2019chemical,krasnopolsky2020photochemical} that the formation of hydrocarbon-based hazes starts with the photolysis of ``haze precursor'' molecules such as \ce{CH4}, \ce{C2H2}, HCN and \ce{C6H6}. Except for the \ce{C6H6}, all of these species are included in our reaction network. The abundances of these species can thus give a rough sense of the importance of photochemical hazes. Using a combination of a kinetics model and an aerosol microphysics model, \citet{lavvas2017aerosol} found that hazes can be efficiently formed on \revisionone{some of the colder} hot-jupiters, such as HD\;189733b \revisionone{with an equilibrium temperature of $\sim$1100\;K}. On the hotter HD\;209458b, \revisionone{T\textsubscript{eq}$\approx$1300\;K}, photochemical hazes did not form efficiently enough to affect the predicted transit spectra. Comparing the dayside abundances of CH$_4$, C$_2$H$_2$ and HCN from our kinetic model, we find that the abundances of each of these haze precursor molecules are either comparable to or significantly lower than the abundances in the HD\;209458b simulation. We thus conclude that photochemical hazes are even less important for HAT-P-7b. Furthermore, temperatures exceeding 2700\;K in the upper atmosphere on large parts of the day side preclude even hazes with properties similar to soot particles in these regions, as they are not stable at these temperatures \citep{rimmer2013cosmic}.


\section{Conclusions} \label{s:conclusions}
In this paper we investigated the importance of chemical disequilibrium (namely vertical mixing and photochemistry) on an ultra-hot Jupiter, HAT-P-7b, as an example of this class of planets.

We employed two approaches to investigate the quenching levels: a zeroth-order approximation based on \citet{venot2018better} work and full kinetic modeling. While both approaches usually agree with each other, the zeroth-order approximation underestimates the quenching levels at high temperatures by about one order of magnitude. 

By performing \revisionone{1D chemical kinetic simulations for 38 molecules and at 97 latitude and longitude points on a 3D GCM model,} we find that the disequilibrium is a local phenomena and affects the nightside and the morning terminator the most for specific molecules like H, \ce{H2O}, \ce{CH4}, \ce{CO2}, \ce{HCN}, and all C$_n$H$_m$ molecules; heavier C$_n$H$_m$ molecules are more affected by disequilibrium processes. For those molecules, the variation of abundances on the dayside is also noticeable but the evening terminator is the least affected region by disequilibrium processes. This may partially explain the observed transmission spectra of ultra-hot Jupiters to be consistent with thermochemical equilibrium state.

We find that the CO abundance is affected by vertical mixing only marginally, whereas the abundance of other major opacity species such as \ce{CH4}, \ce{H2O}, and also HCN are influenced by these processes. Photochemistry has almost no effect on the abundance of species considered in this study.

Thanks to the locality characteristic of disequilibrium processes, we propose four avenues to look for the effect of these processes on ultra-hot Jupiters. All four methods are based on comparative retrieval, where local atmospheric properties are retrieved and compared. We, however, caution that clouds will have to be taken into account for any interpretation. In the first method we suggest to investigate vertical opacity distributions between the morning terminator (in transmission) and the sub-stellar point (in emission). The second method is to look for differences observed in transmission spectra between the morning and evening terminators. The third method is to measure C/O ratio at the two terminators and look for a lack of asymmetry (C/O ratio could deviate by 0.2 between the two limbs if only cloud formation is considered). A similar C/O ratio at the two terminators could indicate that the disequilibrium processes are in action. An enhancement of global observed C/O ratio (by about 0.05 in the case of HAT-P-7b) could also be an indication of cloud formation on this planet. Given the challenging nature of these methods, probably the best chance to observe the effect of clouds and disequilibrium chemistry with the current facilities is using precise phase curves \revisionone{as the fourth avenue}. Through such precision photometry, a longitudinal chemical map of the planet could be constructed as reported by \citet{tan2019atmospheric,arcangeli2019climate,mansfield2019evidence,wong2019exploring}. Finally, it is worth mentioning that the 3D atmospheric structure of HAT-p-7b and possible geometric expansion of the dayside over the nightside, as described by \citet{caldas2019effects}, could make the retrieval of these parameters even more challenging. We investigate and discuss the observability of HAT-P-7b's atmosphere and the effects of disequilibrium on it in details in \revisionone{a forthcoming paper}.


\begin{acknowledgements}
We acknowledge Cloud Academy I 2018 at the Les Houches School of Physics, France, during which this project was kicked off. We thank Maria Steinrueck for the constructive  discussions and contribution. Part of this work was supported by the German \emph{Deut\-sche For\-schungs\-ge\-mein\-schaft, DFG\/} project number Ts~17/2--1.
\end{acknowledgements}

\appendix
\section{Supplemental Figures} \label{s:appendix}
Additional figures are provided in this section.

\begin{figure*}
    \includegraphics[width=0.45\textwidth]{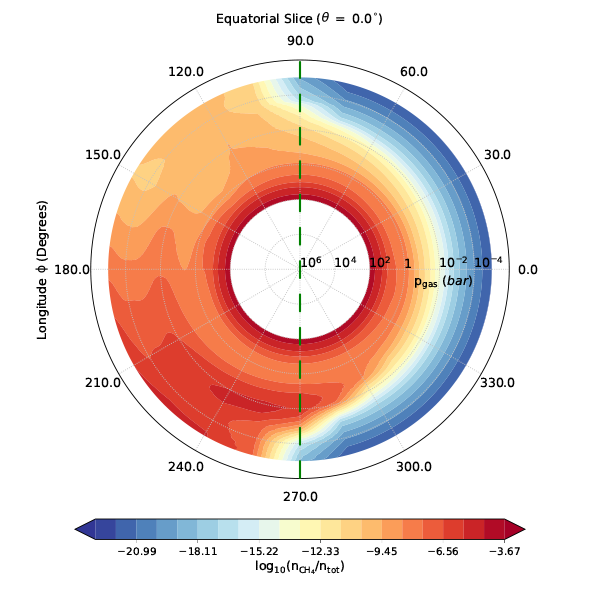}
        \includegraphics[width=0.45\textwidth]{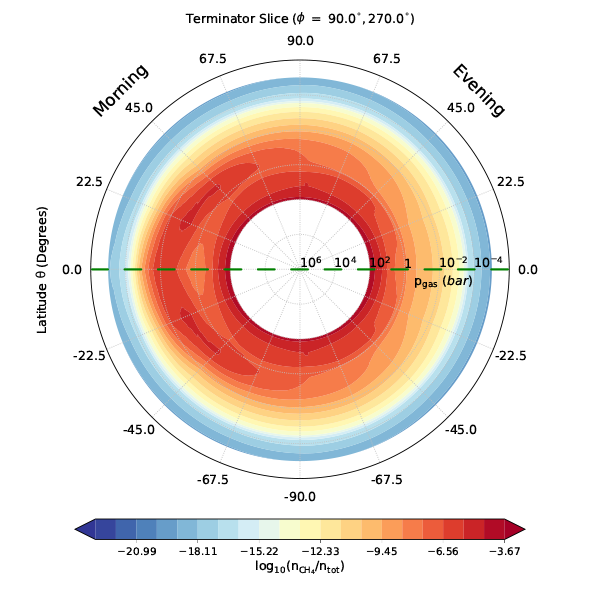}\\*[-0.4cm]
    \includegraphics[width=0.45\textwidth]{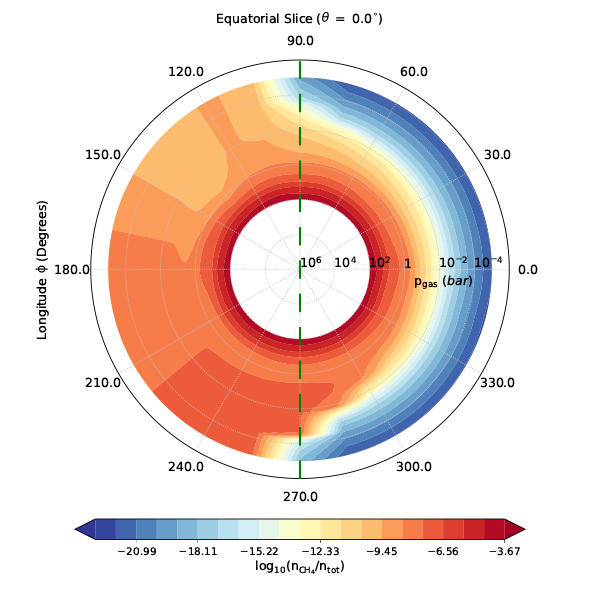}
        \includegraphics[width=0.45\textwidth]{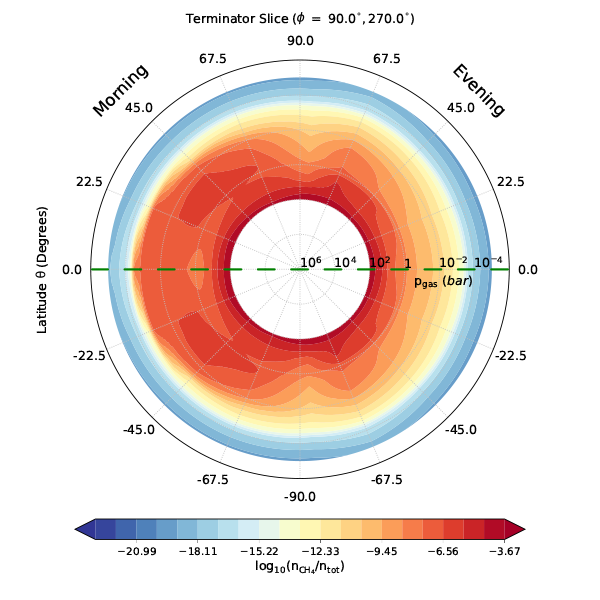}\\*[-0.4cm]
    \includegraphics[width=0.45\textwidth]{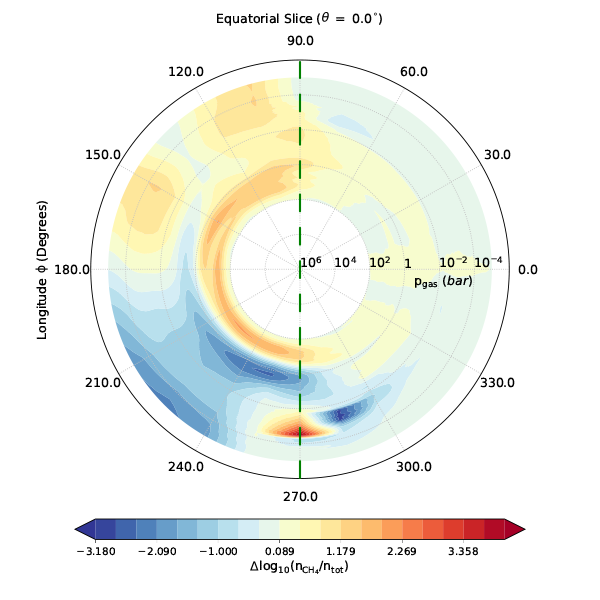}
       \includegraphics[width=0.45\textwidth]{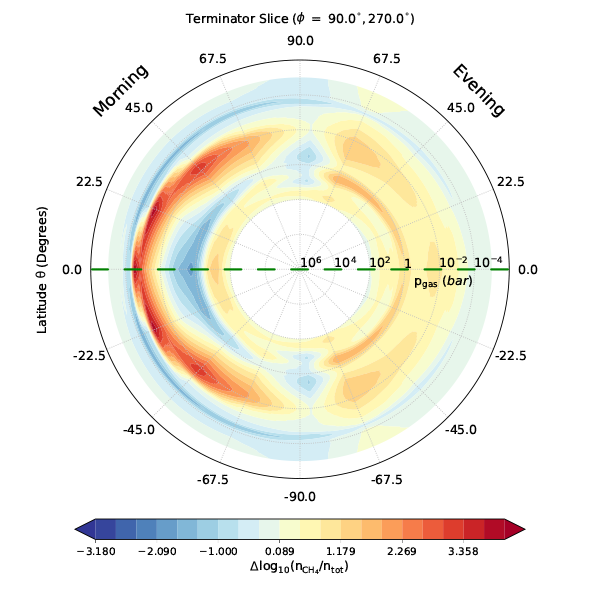}\\*[-1cm]
    \caption{Methane (\ce{CH4}):  {\bf Left column:} equatorial slices ($\theta=0\degree$), {\bf Right column:} terminator slices; 
    \textbf{Top:} initial condition, $\log(n_{\rm \ce{CH4}}^{\rm init}/n_{\rm tot})$; \textbf{Middle:}  final result from kinetic simulation, $\log(n_{\rm \ce{CH4}}^{\rm final}/n_{\rm tot})$ \textbf{Bottom:} difference, $\log(n_{\rm \ce{CH4}}^{\rm final}/n_{\rm init})$ between the initial (equilibrium) values and the kinetic results. 
}
    \label{fig:CH4_results}
\end{figure*}

\begin{figure*}
    \includegraphics[width=0.45\textwidth]{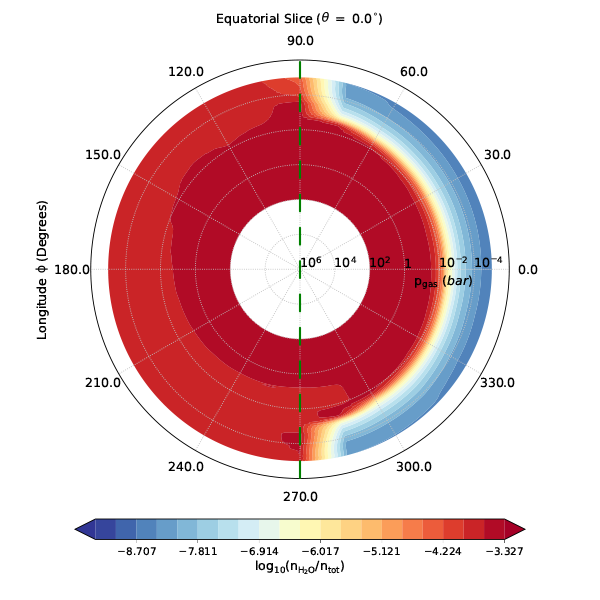}
    \includegraphics[width=0.45\textwidth]{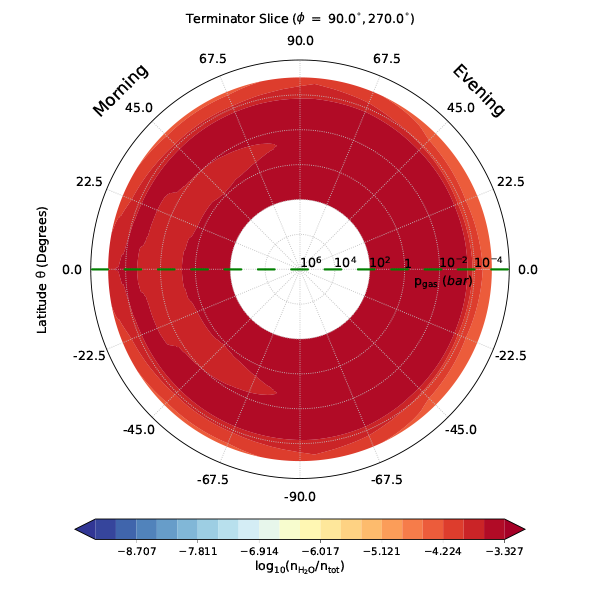}\\*[-0.4cm]
    \includegraphics[width=0.45\textwidth]{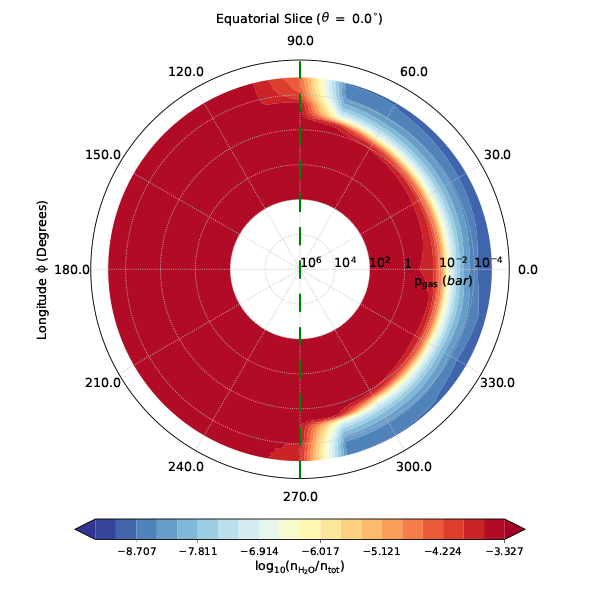}
    \includegraphics[width=0.45\textwidth]{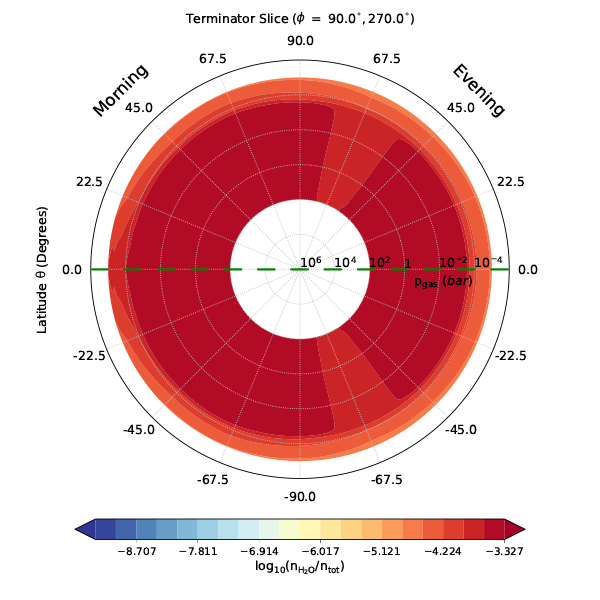}\\*[-0.4cm]
    \includegraphics[width=0.45\textwidth]{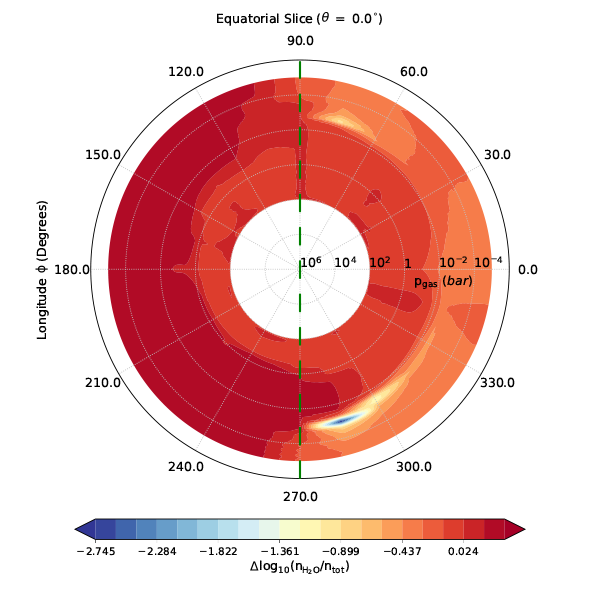}
    \includegraphics[width=0.45\textwidth]{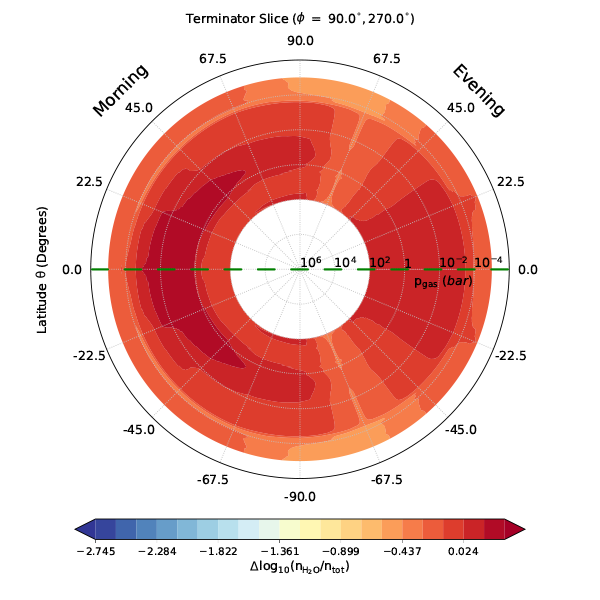}\\*[-1cm]
    \caption{Water (\ce{H2O}):  {\bf Left column:} equatorial slices ($\theta=0\degree$), {\bf Right column:} terminator slices; 
    \textbf{Top:} initial condition, $\log(n_{\rm \ce{H2O}}^{\rm init}/n_{\rm tot})$; \textbf{Middle:}  final result from kinetic simulation, $\log(n_{\rm \ce{H2O}}^{\rm final}/n_{\rm tot})$ \textbf{Bottom:} difference, $\log(n_{\rm \ce{H2O}}^{\rm final}/n_{\rm init})$ between the initial (equilibrium) values and the kinetic results.}
    \label{fig:H2O_results}
\end{figure*}

\begin{figure*}
    \includegraphics[width=0.45\textwidth]{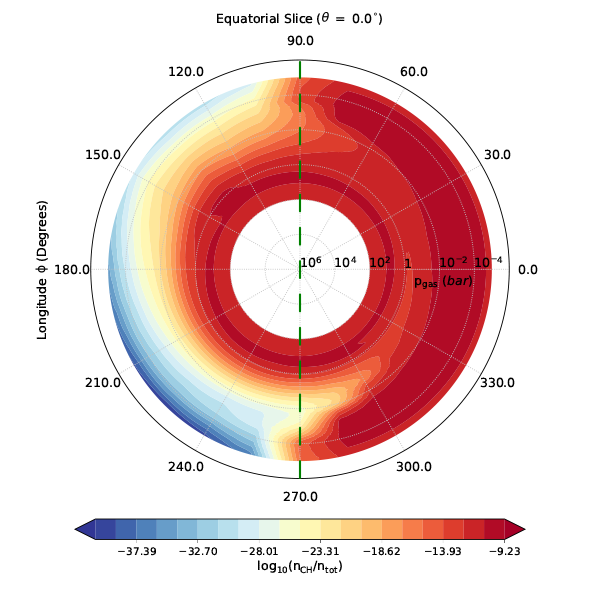}
        \includegraphics[width=0.45\textwidth]{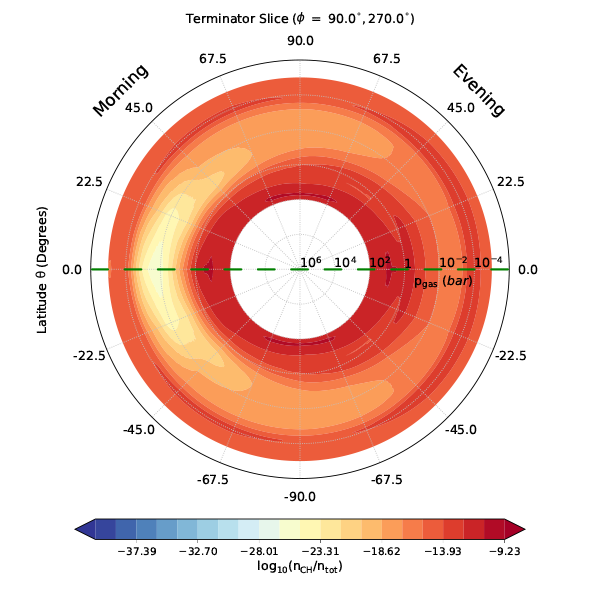}\\*[-0.4cm]
    \includegraphics[width=0.45\textwidth]{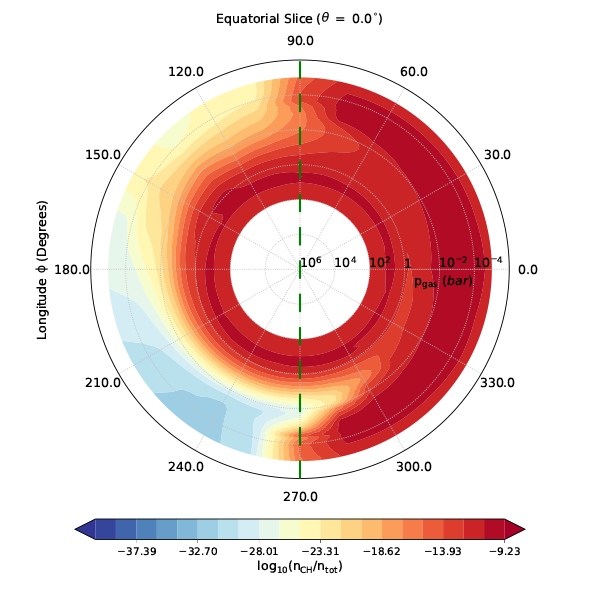}
        \includegraphics[width=0.45\textwidth]{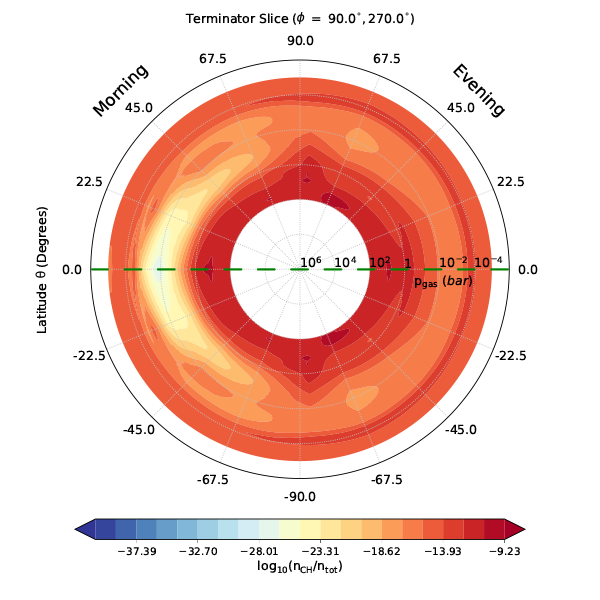}\\*[-0.4cm]
    \includegraphics[width=0.45\textwidth]{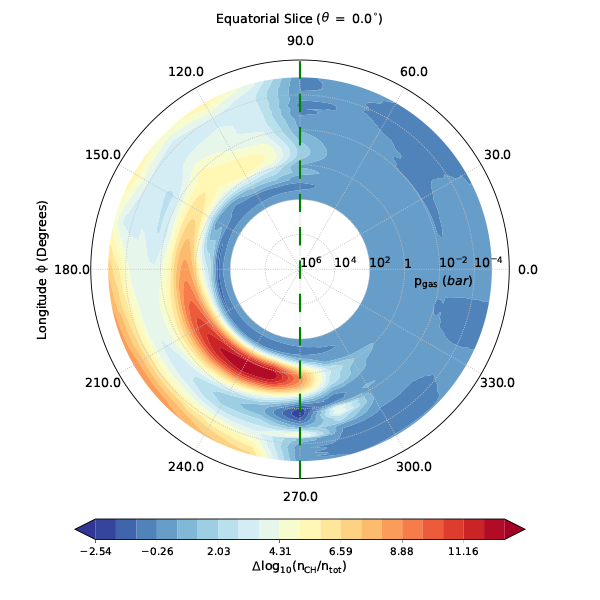}
       \includegraphics[width=0.45\textwidth]{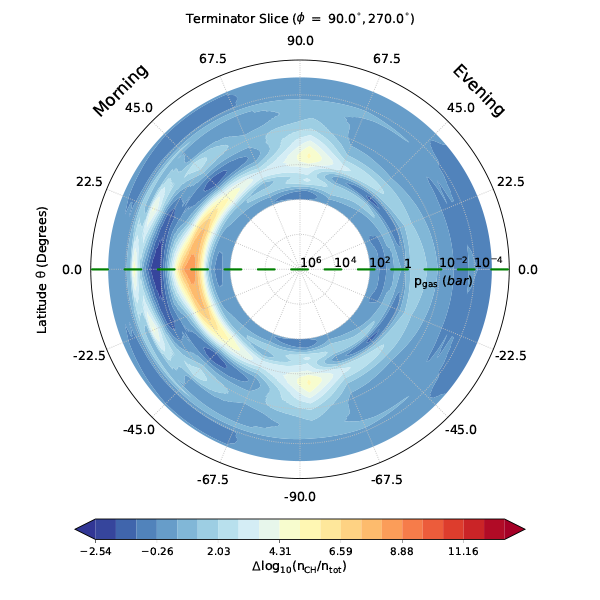}\\*[-1cm]
    \caption{Methylidyne (carbyne) (\ce{CH}):  {\bf Left column:} equatorial slices ($\theta=0\degree$), {\bf Right column:} terminator slices; 
    \textbf{Top:} initial condition, $\log(n_{\rm CH}^{\rm init}/n_{\rm tot})$; \textbf{Middle:}  final result from kinetic simulation, $\log(n_{\rm CH}^{\rm final}/n_{\rm tot})$ \textbf{Bottom:} difference, $\log(n_{\rm CH}^{\rm final}/n_{\rm init})$ between the initial (equilibrium) values and the kinetic results. 
}
    \label{fig:CH_results}
\end{figure*}

\begin{figure*}
    \includegraphics[width=0.45\textwidth]{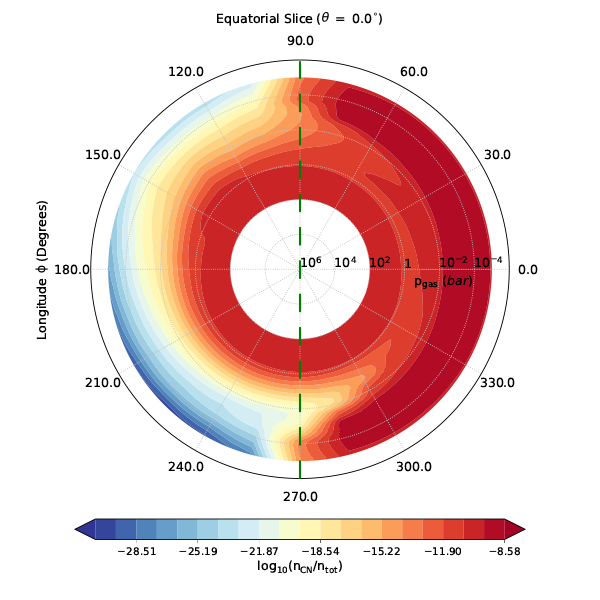}
    \includegraphics[width=0.45\textwidth]{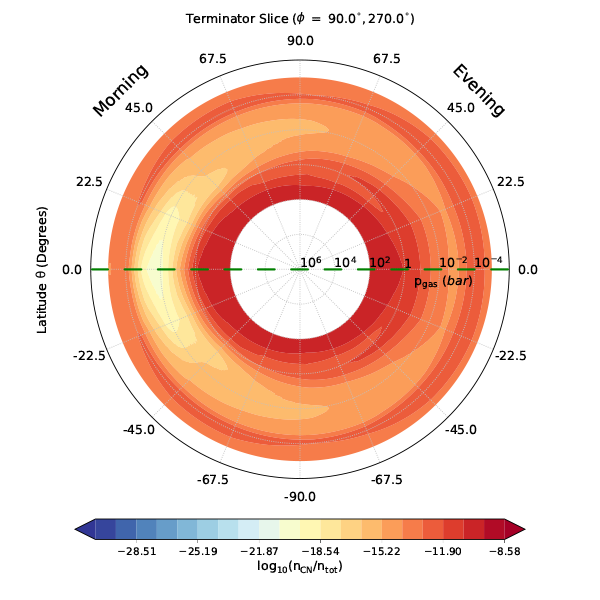}\\*[-0.4cm]
    \includegraphics[width=0.45\textwidth]{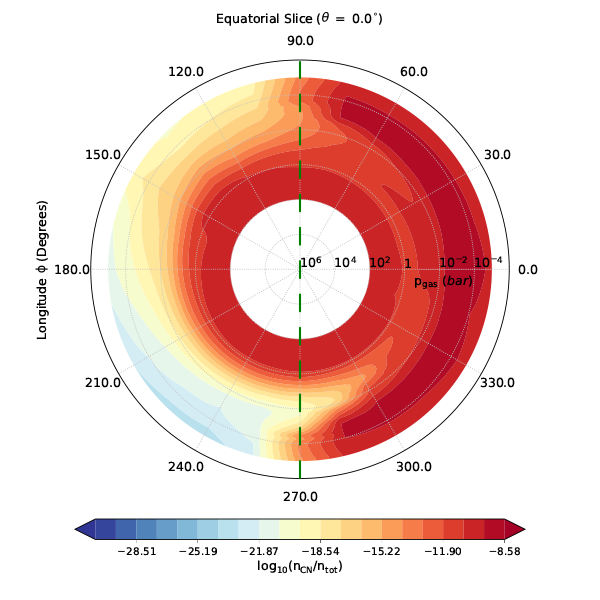}
    \includegraphics[width=0.45\textwidth]{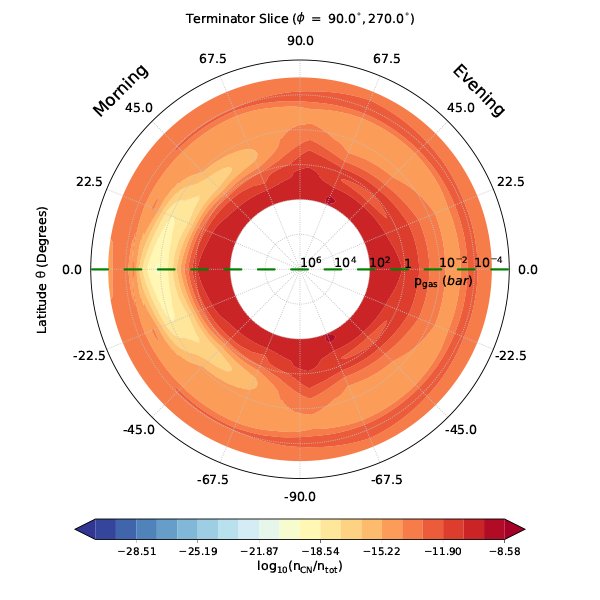}\\*[-0.4cm]
    \includegraphics[width=0.45\textwidth]{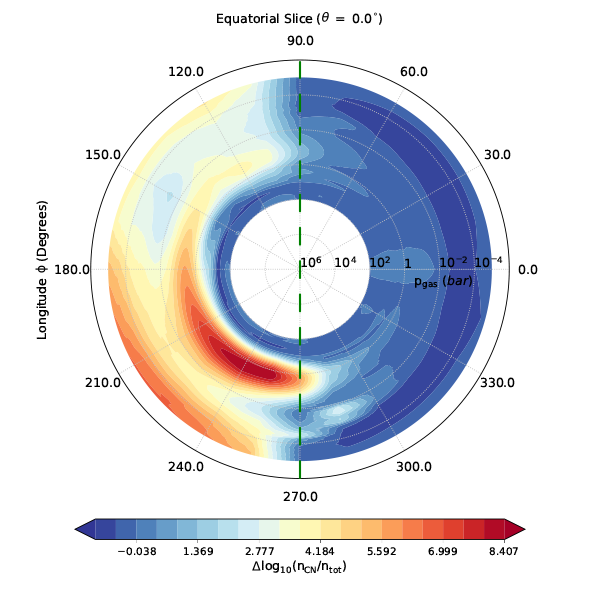}
    \includegraphics[width=0.45\textwidth]{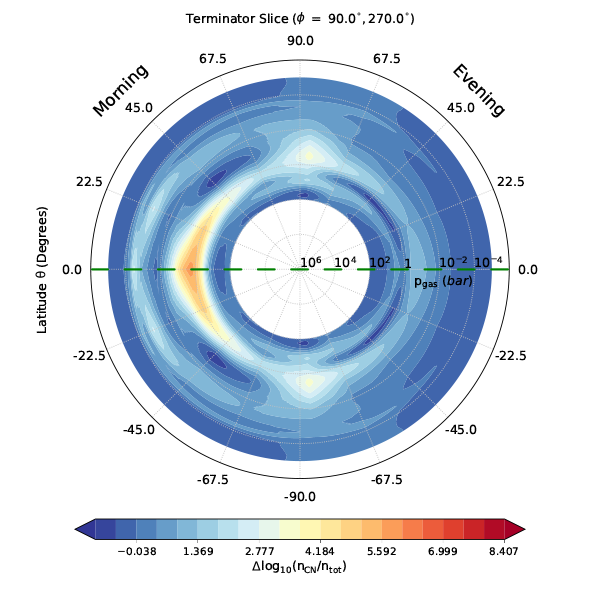}\\*[-1cm]
    \caption{Cyanide (\ce{CN}):  {\bf Left column:} equatorial slices ($\theta=0\degree$), {\bf Right column:} terminator slices; 
    \textbf{Top:} initial condition, $\log(n_{\rm CN}^{\rm init}/n_{\rm tot})$; \textbf{Middle:}  final result from kinetic simulation, $\log(n_{\rm CN}^{\rm final}/n_{\rm tot})$ \textbf{Bottom:} difference, $\log(n_{\rm CN}^{\rm final}/n_{\rm init})$ between the initial (equilibrium) values and the kinetic results. 
}
    \label{fig:CN_results}
\end{figure*}

\begin{figure*}
    \includegraphics[width=0.45\textwidth]{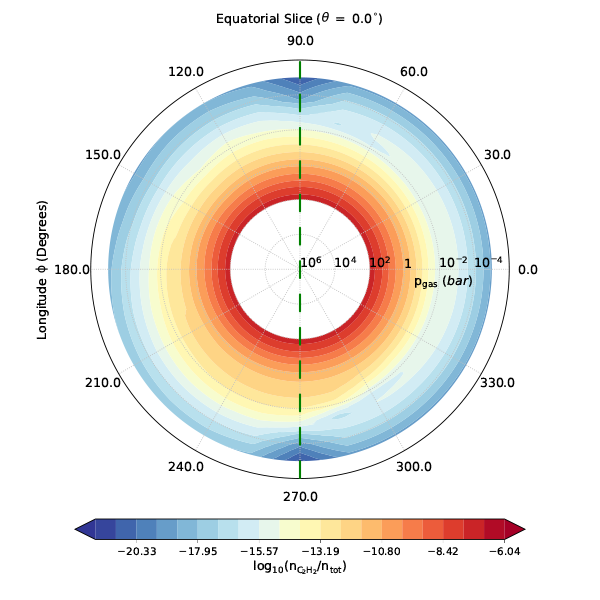}
        \includegraphics[width=0.45\textwidth]{Diffusion_Photochem/Diffusion_Photochem_Startend_CN_5.png}\\*[-0.4cm]
    \includegraphics[width=0.45\textwidth]{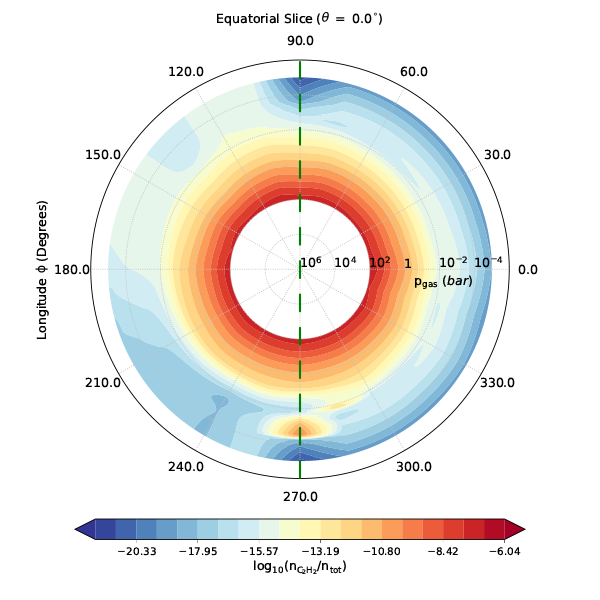}
        \includegraphics[width=0.45\textwidth]{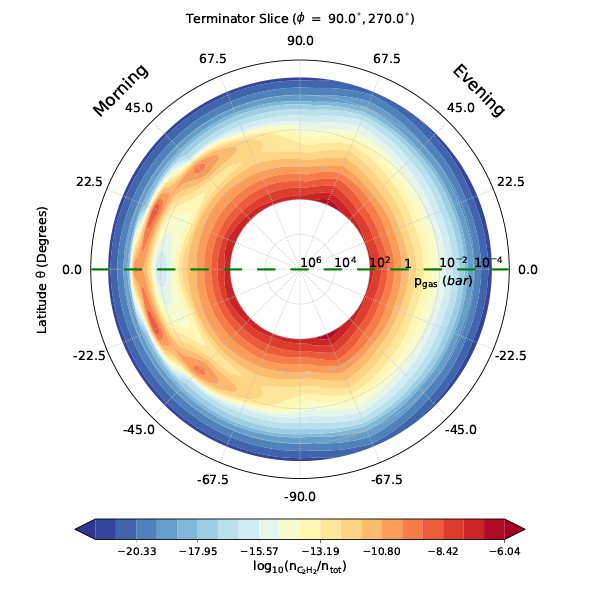}\\*[-0.4cm]
    \includegraphics[width=0.45\textwidth]{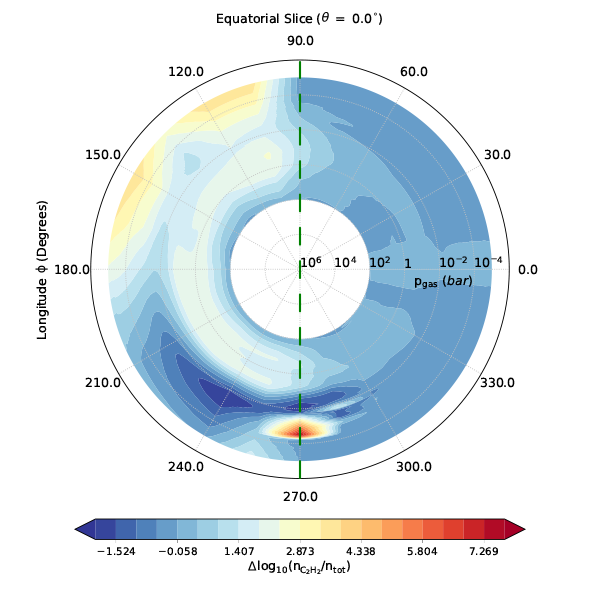}
       \includegraphics[width=0.45\textwidth]{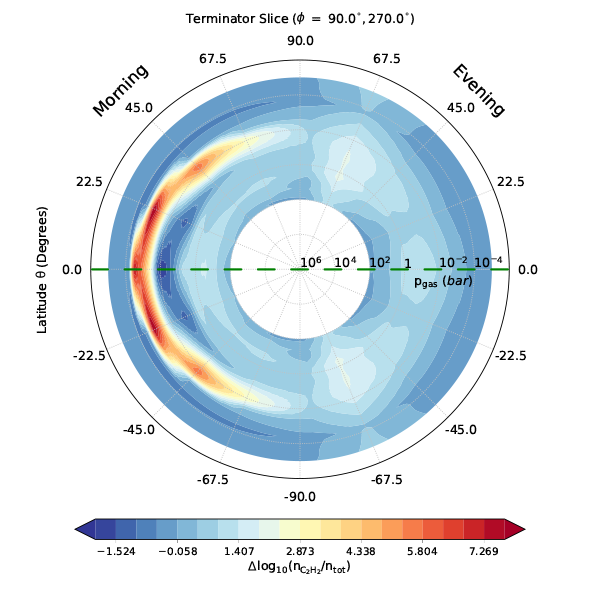}\\*[-1cm]
    \caption{Acetylene (\ce{C2H2}):  {\bf Left column:} equatorial slices ($\theta=0\degree$), {\bf Right column:} terminator slices; 
    \textbf{Top:} initial condition, $\log(n_{\rm C2H2}^{\rm init}/n_{\rm tot})$; \textbf{Middle:}  final result from kinetic simulation, $\log(n_{\rm C2H2}^{\rm final}/n_{\rm tot})$ \textbf{Bottom:} difference, $\log(n_{\rm C2H2}^{\rm final}/n_{\rm init})$ between the initial (equilibrium) values and the kinetic results. 
}
    \label{fig:C2H2_results}
\end{figure*}

\begin{figure*}
    \includegraphics[width=0.45\textwidth]{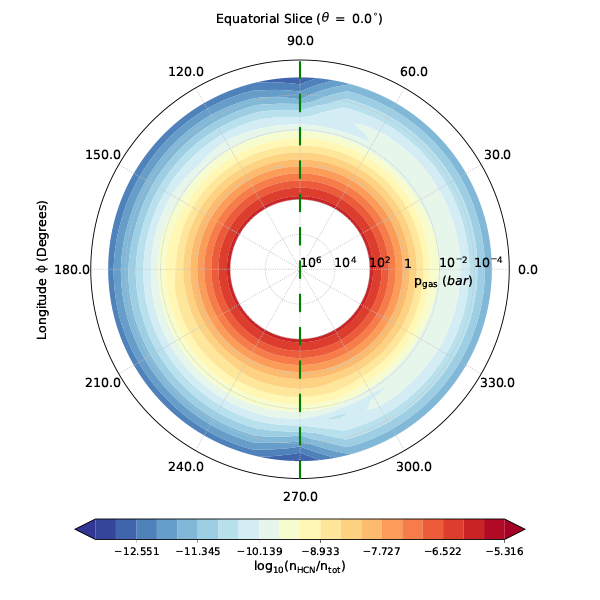}
    \includegraphics[width=0.45\textwidth]{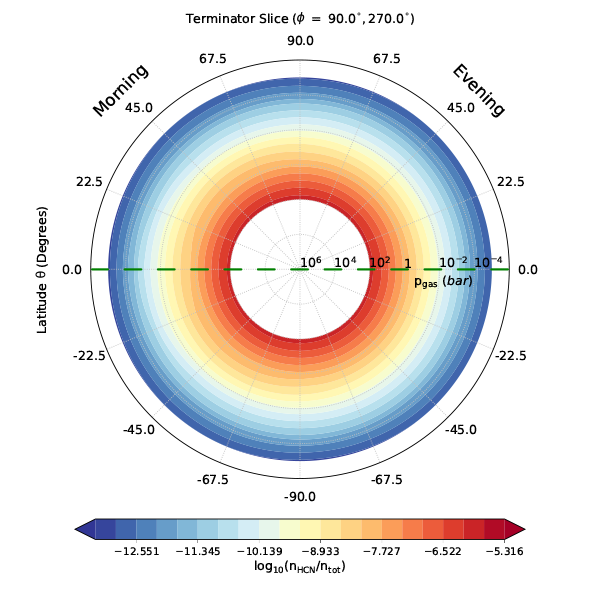}\\*[-0.4cm]
    \includegraphics[width=0.45\textwidth]{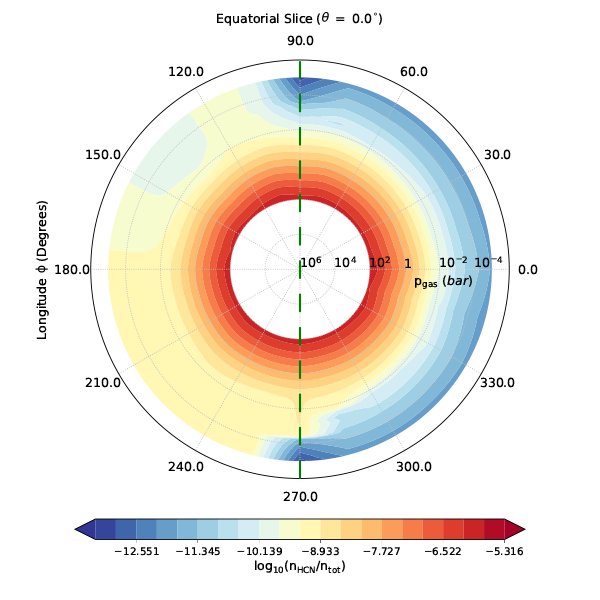}
    \includegraphics[width=0.45\textwidth]{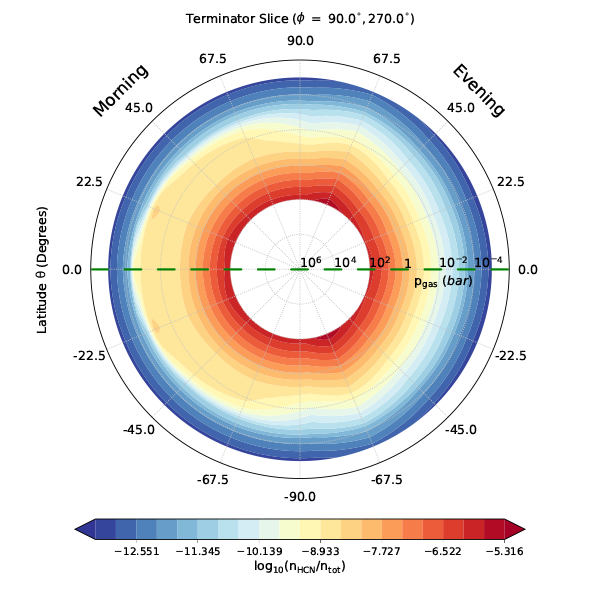}\\*[-0.4cm]
    \includegraphics[width=0.45\textwidth]{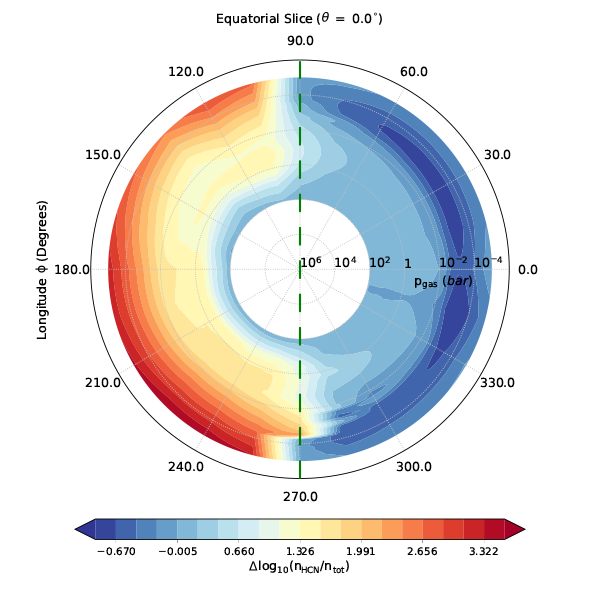}
    \includegraphics[width=0.45\textwidth]{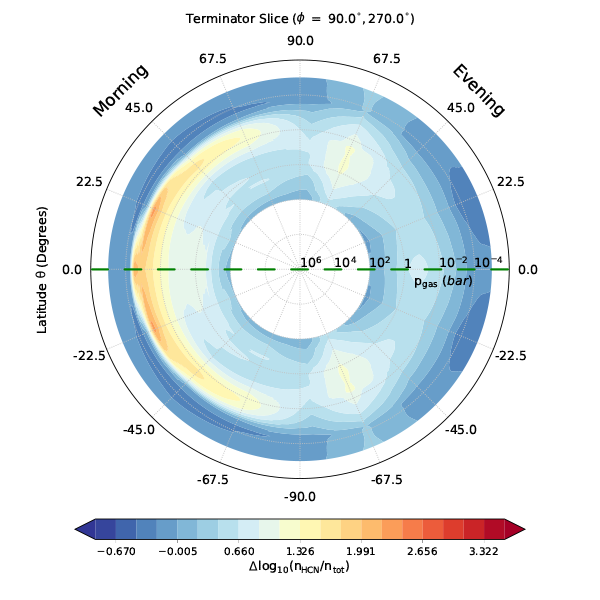}\\*[-1cm]
    \caption{Hydrogen cyanide (\ce{HCN}):  {\bf Left column:} equatorial slices ($\theta=0\degree$), {\bf Right column:} terminator slices; 
    \textbf{Top:} initial condition, $\log(n_{\rm HCN}^{\rm init}/n_{\rm tot})$; \textbf{Middle:}  final result from kinetic simulation, $\log(n_{\rm HCN}^{\rm final}/n_{\rm tot})$ \textbf{Bottom:} difference, $\log(n_{\rm HCN}^{\rm final}/n_{\rm init})$ between the initial (equilibrium) values and the kinetic results.}
    \label{fig:HCN_results}
\end{figure*}

\begin{figure*}[h]
\includegraphics[width=\textwidth]{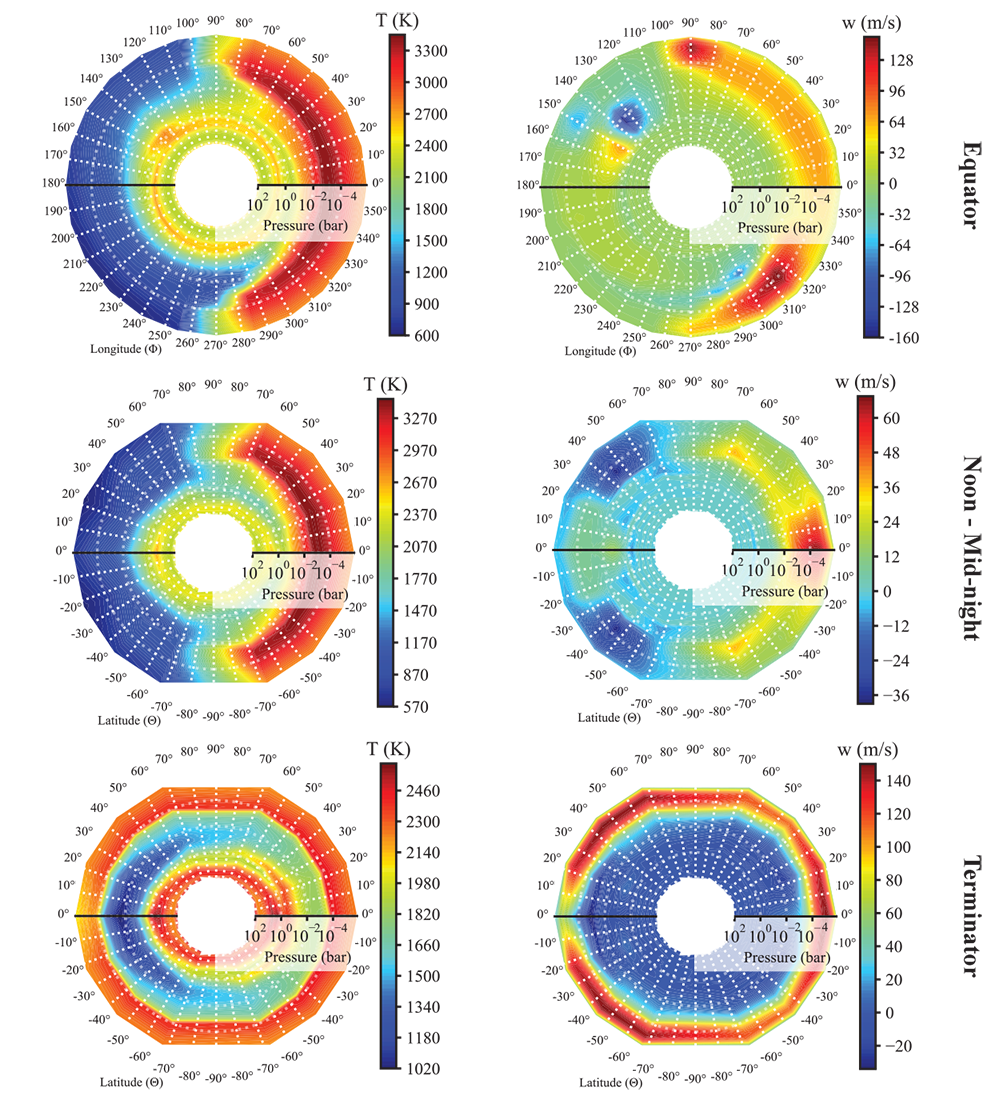}
\caption{Transection maps of temperature (K) and vertical velocities (w) obtained from the GCM solution.}
\label{fig:quench0th maps}
\end{figure*}

\begin{figure*}
    \includegraphics[width=0.45\textwidth]{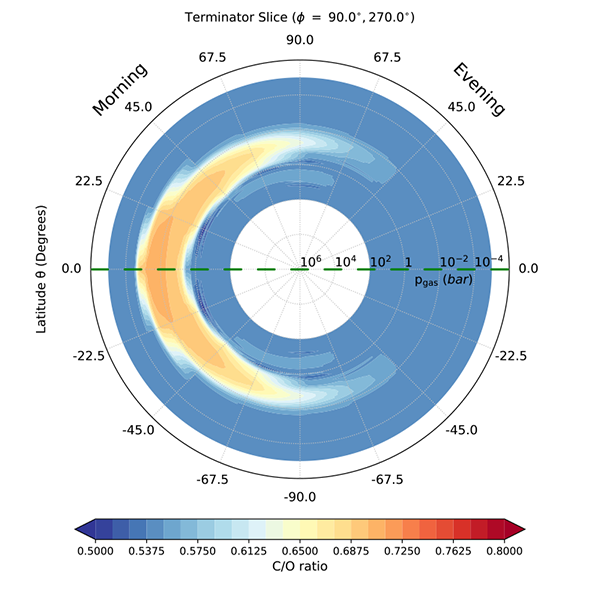}
    \includegraphics[width=0.45\textwidth]{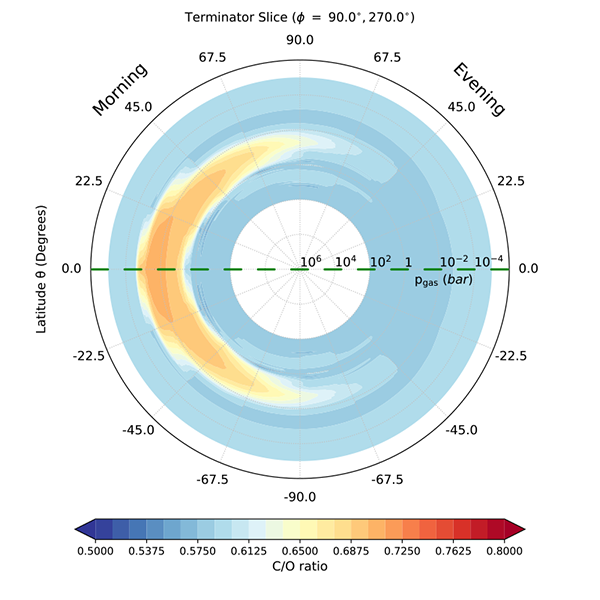}\\
    \includegraphics[width=0.45\textwidth]{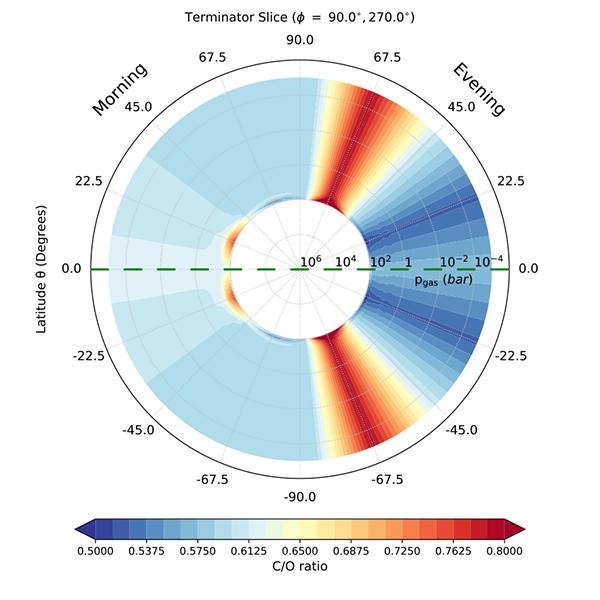}
    \includegraphics[width=0.45\textwidth]{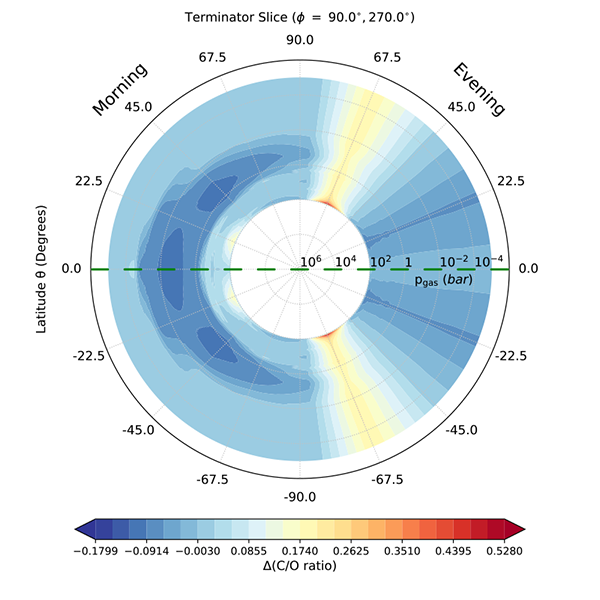}
    \caption{The carbon-to-oxygen ratio (C/O) in terminator slices representation: \textbf{top left:} C/O after cloud formation as in Paper I according to $\epsilon_{C}/\epsilon_{O}$, 
    \textbf{top right:} C/O for initial condition of gas-kinetic calculations (from CH, CO, \ce{CO2}, \ce{C2H2, ...), },  
    \textbf{bottom left:} C/O at end of gas-kinetic calculation (t=300yrs), \textbf{bottom right:} C/O change between initial condition (top right) and final gas-kinetic solution (bottom left).
    See appendix Fig.~\ref{fig:CO_equ_Ratio_TEST} for terminator slice representation.
}
    \label{fig:CO_equ_Ratio_terminator}
\end{figure*}

\section{Approximation of Molecular Diffusion Coefficients} \label{s:appendix_mol_diff}
We described our chemical kinetic modelling in Section \ref{sec:kineticmodelling}. Here we expand on how we approximate the molecular diffusion coefficients using the Chapman and Enskog equation, Eq. \ref{eq:diff}, \citep{poling_properties_2000} -- recall:
\begin{equation}
D=\frac{3}{16}\frac{(4\pi kT/M_{AB})^{1/2}}{n\pi \sigma^2_{AB}\Omega_{D}}f_D
\end{equation}
where $M_{AB}$ is defined as $2[(1/M_A)+(1/M_B)]^{-1}$ ($M_A$, $M_B$ being the molecular weights of species A and B, respectively), $\Omega_{D}$ is the collision integral for diffusion, $\sigma_{AB}$ is the characteristic length of the intermolecular force law, $f_D$ is a correction term (usually of order unity), n is number density of molecules in the mixture, k is the Boltzmann’s constant, and T is the gas temperature of the mixture.Among them, $\Omega_{D}$ and $\sigma_{AB}$ demand extra steps to be approximated.

$\Omega_{D}$ can be estimated by following the relation of \citet{neufeld1972empirical}:
\begin{equation}
\Omega_D=\frac{A}{(T^*)^B}+\frac{C}{exp(DT^*)}+\frac{E}{exp(FT^*)}+\frac{G}{exp(HT^*)}
\end{equation}
where A=1.06036, B=0.15610, C=0.19300, D=0.47635, E=1.03587, F=1.52996, G=1.76474, H=3.89411, and $T^*$=$kT/\varepsilon_{AB}$. $\varepsilon_{AB}$ can be estimated as:
\begin{equation}
\varepsilon_{AB}=(\varepsilon_{A}\varepsilon_{B})^{1/2}
\end{equation}
where $\varepsilon_{A}$ and $\varepsilon_{B}$ are the characteristic Lennard-Jones energy of molecule A and B respectively, which their values are usually tabulated and in-hand \citep[see e.g.][Appendix B]{poling_properties_2000}.

Similarly, $\sigma_{AB}$ can be estimated as below:
\begin{equation}
\sigma_{AB}=\frac{\sigma_{A}+\sigma_{B}}{2}
\end{equation}
where $\sigma_{A}$ and $\sigma_{B}$ are the characteristic Lennard-Jones length of molecule A and B respectively, which their values are also tabulated \citep[see e.g.][Appendix B]{poling_properties_2000}. Using these approximations the molecular diffusion coefficients can be estimated in the chemical kinetic models.

\bibliographystyle{aa}
\bibliography{reference.bib}

\end{document}